\documentclass[conference]{IEEEtran}
\IEEEoverridecommandlockouts
\usepackage{cite}
\usepackage{capt-of}
\usepackage{amsmath,amssymb,amsfonts}
\usepackage{algorithmic}
\usepackage{graphicx}
\usepackage{textcomp}
\usepackage{xcolor}
\usepackage{multirow}
\usepackage{graphicx}
\usepackage{subcaption} 
\usepackage{adjustbox}   % Helps in centering subfigures
\usepackage{verbatim}
\usepackage{fancyvrb}
\usepackage{url}
\usepackage{tikz}
\usetikzlibrary{arrows.meta, positioning, shapes, calc, fit}
\usepackage{pifont}% http://ctan.org/pkg/pifont
\newcommand{\xmark}{\ding{55}}%
\usepackage[most]{tcolorbox}
\usepackage{float} % For floating box with !t
\usepackage{placeins}
\usepackage{tabularx}
\tcbset{
  ddossummary/.style={
    floatplacement=!ht,
    float,
    colback=red!5!white,
    colframe=black,
    boxrule=0.5pt,
    arc=4pt,
    left=6pt,
    right=6pt,
    top=6pt,
    bottom=6pt,
    enhanced,
    sharp corners=south,
    before skip=10pt,
    after skip=10pt,
    title={\textbf{DDoS Detection Model Output Summary}},
  }
}

\def\BibTeX{{\rm B\kern-.05em{\sc i\kern-.025em b}\kern-.08em
    T\kern-.1667em\lower.7ex\hbox{E}\kern-.125emX}}
\begin{document}

\title{Interpretable Anomaly-Based DDoS Detection in AI-RAN with XAI and LLMs}

\author{
\IEEEauthorblockN{
    Sotiris Chatzimiltis\IEEEauthorrefmark{1}, \textit{Graduate Student Member, IEEE}, 
    Mohammad Shojafar\IEEEauthorrefmark{1}, \textit{Senior Member, IEEE}, \\
    Mahdi Boloursaz Mashhadi\IEEEauthorrefmark{1}, \textit{Senior Member, IEEE}, 
    and Rahim Tafazolli\IEEEauthorrefmark{1}, \textit{Fellow, IEEE}
}
\IEEEauthorblockA{
    \IEEEauthorrefmark{1}5G/6GIC, Institute for Communication Systems (ICS), University of Surrey, Guildford, UK \\
    \{sc02449, m.shojafar, m.boloursazmashhadi, r.tafazolli\}@surrey.ac.uk
}
}
\maketitle

\begin{abstract}
Next generation Radio Access Networks (RANs) introduce programmability, intelligence, and near real-time control through intelligent controllers, enabling enhanced security within the RAN and across broader 5G/6G infrastructures. This paper presents a comprehensive survey highlighting opportunities, challenges, and research gaps for Large Language Models (LLMs)-assisted explainable (XAI) intrusion detection (IDS) for secure future RAN environments. Motivated by this, we propose an LLM interpretable anomaly-based detection system for distributed denial-of-service (DDoS) attacks using multivariate time series key performance measures (KPMs), extracted from E2 nodes, within the Near Real-Time RAN Intelligent Controller (Near-RT RIC). An LSTM-based model is trained to identify malicious User Equipment (UE) behavior based on these KPMs. To enhance transparency, we apply post-hoc local explainability methods such as LIME and SHAP to interpret individual predictions. Furthermore, LLMs are employed to convert technical explanations into natural-language insights accessible to non-expert users. Experimental results on real 5G network KPMs demonstrate that our framework achieves high detection accuracy (F1-score $>$ 0.96) while delivering actionable and interpretable outputs. 
\end{abstract}


\section{Introduction}\label{sec:introduction}
The Open Radio Access Network (RAN) paradigm has significantly reshaped the modern wireless network architecture by introducing disaggregated, interoperable, and programmable components into the RAN~\cite{oranwhitepaper2018,polese2023understanding}. A key enabler of this flexibility is the introduction of intelligent RAN controllers, specifically the Near-Real-Time RIC (Near-RT RIC) and the Non-Real-Time RIC (Non-RT RIC), which respectively support the deployment of custom applications known as xApps and rApps~\cite{balasubramanian2021ric,soltani2025ric}. These applications facilitate intelligent control, optimization, and monitoring of radio resources by leveraging analytics and machine learning (ML) across different time scales. This software-driven programmability enhances network adaptability and automation, supporting dynamic network slicing, advanced security mechanisms, resource allocation, and orchestration processes that flexibly adapt to diverse and evolving network demands~\cite{ojaghi2022slicedRAN,alemany2025Intent6G}.
Building on these architectural foundations, the AI RAN paradigm has emerged as a natural next step, aiming to integrate AI-driven intelligence across all layers of the RAN. Defined by the AI-RAN Alliance, this paradigm outlines three key development areas: AI-for-RAN, AI-on-RAN, and AI-and-RAN~\cite{ai-ran-whitepaper,kundu2025airantransformingranaidriven}, each addressing a distinct aspect of AI integration in the design and operation of RAN systems.

In this context, AI RAN can provide opportunities to harness AI to improve network security. Both AI-for-RAN and AI-on-RAN enable the development of intelligent, adaptive mechanisms for detecting and mitigating threats. This is particularly critical as mobile networks increasingly face complex and evolving security challenges.
Among the most critical threats to any mobile network infrastructure are distributed denial-of-service (DDoS) attacks~\cite{donnghyun6gthreats}, often initiated by compromised or malicious user equipment (UE). These attacks can overwhelm the control or data planes of the network, leading to service degradation and potential outages~\cite{5gcoresec2022}. While traditional DDoS defenses often reside in the core network, detecting anomalous behavior early, closer to the edge, can significantly reduce response time and damage. Leveraging telemetry data, such as key performance measurements (KPMs) exposed via E2 interfaces~\cite{oran2025e2smkpm},\cite{oran2025e2gap}, \cite{oran2025e2ap}, the Near-RT RIC can host intelligent DDoS detection mechanisms as xApps that analyze UE behavior in near real time and enable timely detection. This work contributes to the AI-for-RAN development area by proposing an ML-based framework for detecting DDoS attacks using multivariate time-series KPMs collected from UEs. A long short-term memory (LSTM) model is trained to classify UE behavior as benign or malicious based on these temporal features. 

Although detection models can achieve strong predictive performance, their decision process remains opaque, which poses challenges for trust, validation, and operational use in security-critical environments~\cite{subashIDSxai,alfredoXAIsurvey}.
This is where the AI-on-RAN paradigm plays a key role, by enabling the deployment of auxiliary AI models, including large language models (LLMs) and generative AI tools. Recent advancements in AI-native network architectures~\cite{Chen2024NetGPT, 10742565, 10614204} exemplify this approach, showcasing the collaborative deployment of LLMs across cloud and edge environments. Such architectures facilitate real-time generative AI services and intelligent network management that can enhance RAN functionality, improve overall network security, and support operator decision-making.

Existing works either focus solely on detection or explainability without integrating automated reasoning via LLMs, or they do not explicitly address the operational time constraints critical for deployment in Near-RT RIC environments.
To address this, our framework first integrates Explainable AI (XAI) techniques to provide local explanations for individual LSTM predictions, highlighting which features contributed to a specific classification. However, raw XAI output, such as feature attribution scores, can be complex and difficult for non-expert personnel to interpret quickly and act upon.
To overcome this, these local explanations are passed through an LLM that automatically translates complex technical outputs into human-readable natural language insights, and it leverages its reasoning capabilities (if applicable) to suggest potential mitigation strategies based on detected anomalies. By doing so, the LLM reduces the cognitive burden on operators, accelerates incident response, and moves toward partial security workflow automation, bridging the gap between anomaly detection and mitigation.

\textbf{Contributions.} The main contributions of this paper are summarized as follows:
\begin{enumerate}
    \item We provide a comprehensive survey on LLM and XAI integration for intrusion detection systems (IDS) for RAN security in future networks. We highlight the key challenges and opportunities as well as the existing research gaps in this area. 
    \item We propose an LLM explainable DDoS detection framework leveraging an LSTM-based model operating on UE KPM time-series data.
    \item We integrate local XAI methods (LIME and SHAP) to explain per-instance predictions and enhance interpretability.
    \item We introduce the use of LLMs to translate technical explanations into natural language, increasing human readability.
\end{enumerate}

The remainder of the paper is organized as follows. Section~\ref{sec:rel_works} reviews related work on intrusion detection, XAI, and LLM in wireless networks. Section~\ref{sec:3} introduces the proposed framework, detailing the system architecture, the LSTM-based detection model, and the interpretability pipeline. Section~\ref{sec:4} presents the experimental evaluation, which assesses both detection accuracy and explanation quality. Finally, Section~\ref{sec:5} concludes the paper and outlines potential future directions.

\section{Related Work}\label{sec:rel_works}
This section reviews related work on IDSs, with a focus on the integration of ML and deep learning (DL) techniques, the role of XAI in improving the transparency of IDS decisions, and the emerging use of LLMs for intrusion detection, mitigation, and explainability.

\subsection{Intrusion Detection Systems}
IDSs serve as a critical component in the defense of a multitude of networks. Examples include SDN-based IDSs, which leverage centralized control and network programmability for enhanced threat detection~\cite{ojcoms_chatzimilitis},\cite{chatzimiltisITU},\cite{shaji2024sd}, IoT-oriented IDSs, which address the unique constraints of low-power heterogeneous devices~\cite{hore2023empirical},\cite{sharma2023anomaly},\cite{gueriani2024enhancing}, and IDSs designed for vehicular networks, which address high mobility and real-time communication requirements~\cite{jieIDSVANET},\cite{amuthaIDSVN},\cite{fan2024auto}. Unlike static defenses that rely on predefined rules or known threat signatures, IDSs are designed to dynamically observe network activity and identify potentially harmful patterns. In this subsection, we review key developments in classical, ML-based, and DL-based IDSs, with special attention to their application in mobile networks, 5G / B5G infrastructure, and next-generation RAN architectures.

\subsubsection{Classical, ML-Based and DL-Based IDS}
The evolving landscape of cyber threats and the limitations of rule-based detection systems drove the evolution of IDSs. Early systems relied on signature matching to identify known attacks, but such approaches struggled with novel threats~\cite{iyer2021signatures}. This limitation led to behavior-based and anomaly-driven methods, where deviations from learned normal patterns were flagged as potential intrusions. ML played a central role in enabling this shift by providing data-driven models to analyze complex traffic and user behavior~\cite{sommer2010ids}.
Multiple surveys in the past few years reviewed IDS approaches, each providing a different emphasis on scope, methods, and challenges. 
Ahmad et al.~\cite{ahmad_study_ids} provided a systematic review of network IDS (NIDS) utilizing ML and DL approaches. They highlighted recent methodologies, evaluated their strengths and weaknesses, and identified key datasets and evaluation metrics commonly used in the literature.  
Similarly, Abdulganiyu et al.~\cite{Abdulganiyu_ids_survey} conducted a systematic review of the literature for signature, anomaly, and hybrid-based IDSs, evaluating their performance metrics, datasets used, and attack detection capabilities. 
Xu et al.~\cite{xu2025deeplearningbasedintrusiondetection_survey} focused specifically on the DL-based IDS lifecycle, emphasizing data collection, log parsing, and intrusion detection methods. They thoroughly examined DL methodologies such as graph neural networks (GNNs), transformers, and recurrent models, highlighting their effectiveness in detecting both known and zero-day vulnerabilities. 
Khraisat et al.~\cite{ansam_fl_ids_survey} presented an in-depth survey on the application of federated learning (FL) in IDS, analyzing architectures and aggregation strategies for privacy-preserving intrusion detection. The authors analyzed the strengths and limitations of FL methodologies, addressing significant issues like data privacy, security, and scalability in IDS. 
Lastly, He et al.~\cite{he_advml_ids_survey} reviewed recent literature on adversarial ML, specifically targeting DL-based NIDS. They categorized adversarial attacks into white-box and black-box methods, highlighting their specific applications and limitations in the network security context. Additionally, the authors evaluated existing defensive strategies against adversarial attacks, outlined significant challenges, and proposed future research directions to improve the robustness of DL-based NIDS. 

\subsubsection{Temporal Modeling for IDSs}
In certain intrusion detection scenarios, it is necessary to analyze sequences of temporally ordered data samples rather than isolated instances to accurately identify specific attack patterns. Many behaviors, such as DDoS attacks, become apparent only through temporal dependencies between data points. DL approaches have been shown to be effective in modeling such temporal dependencies for anomaly detection in complex network data~\cite{huang2025advancementAD}. To address this, recent research has incorporated temporal modeling techniques that enable systems to learn from the ordering and evolution of events. This shift has brought increased attention to sequential architectures such as recurrent neural networks (RNNs), LSTMs, gated recurrent units (GRUs), and other more cost-efficient variations~\cite{hua2019LSTM}, and more recently, attention-based and transformer models, which offer improved capabilities for capturing long-range dependencies.
For example, Dash et al.~\cite {dash2025optimized} proposed an optimized LSTM-based framework for anomaly detection, using metaheuristic hyperparameter tuning, to showcase how recurrent models improve detection over conventional ML-based frameworks. Building on these advances, recent studies have shown that integrating attention mechanisms with RNN, CNN, and hybrid CNN-BiLSTM architectures can further enhance temporal feature learning and improve intrusion detection performance~\cite{bhattacharya2024attention, dai2024nidsbilstmattention}. Temporal Convolutional Networks (TCNs) have also been introduced as an effective alternative to modeling long-range dependencies in sequential network traffic~\cite{Nazre_2024}. Lastly, transformer-based architectures have enabled flexible sequence modeling for network intrusion detection and have been applied to optimize the trade-off between detection accuracy and timely detection in real-time settings~\cite{Manocchio_2024,chen2023realtimenetworkintrusiondetection}.

\subsubsection{IDSs in 5G, B5G, and Next-Generation RAN}
The evolution of mobile networks toward 5G, beyond 5G (B5G), and next-generation RANs, including paradigms such as Open RAN and AI-RAN, has introduced substantial new security challenges~\cite{LIYANAGE2023103621, hamroun5gsurvey}. Advanced infrastructures introduce new components, interfaces, and network functions that expand the attack surface. As a result, there is a growing emphasis on AI-driven and ML-based approaches to enhance anomaly detection and intrusion prevention capabilities in both the RAN and  the 5G/B5G core~\cite{amachanghi2024surveyoranids}.

Early efforts in this domain leverage ML techniques for attack detection at the network edge. For example, Xavier et al.~\cite{xavierIDSML} proposed an ML-based IDS xApp that utilizes physical and MAC layer features to detect DoS attacks before they reach the core network, while Amachaghi et al.~\cite{amachaghi_IDS_ORAN_ISCC} developed an ensemble of ML classifiers for anomaly detection. Building on these foundations, subsequent research has explored DL to enhance detection and mitigation capabilities. Kouchaki et al.~\cite{kouchakiIDSSlicing} presented a dual xApp framework in which a self-attention-enhanced RNN autoencoder is used for anomaly detection, and a Secure Slicing xApp orchestrates dynamic network slicing in response to detected threats, thereby integrating security and resource management to improve network resilience.
Multilayer defense strategies have also gained attention. Soleymani et al.~\cite{soleymani2024ddosxapp} proposed a framework that combines a dApp for rapid detection at the RAN level with xApps for in-depth analysis at the near-RT RIC, achieving flexible, layered protection against DDoS attacks. Addressing vulnerabilities specific to the O-RAN architecture, Hung et al.~\cite{xAppE2IDS} proposed an anomaly detector targeting threats originating from the E2 interface and third-party xApps, employing state machine analysis and conformance checking to identify and mitigate unauthorized or malicious signaling.
Finally, the distributed and disaggregated architecture of next-generation RANs has enabled federated learning as a promising approach for privacy-preserving and scalable IDS deployment. Attanayaka et al.~\cite{p2p_fl_openran} introduced a peer-to-peer federated learning framework for anomaly detection in O-RAN, enabling decentralized training and secure parameter aggregation. Rumesh et al.~\cite{rumeshfederated_ORAN_DT} further extended this concept by integrating hierarchical FL within a digital twin environment, allowing for safe and effective model training and validation across multiple network slices.

A variety of ML and DL-based defense mechanisms have been proposed to counter DDoS attacks in 5G and B5G core networks. Sheikhi et al.~\cite{sheikhi2023ddos} introduced an unsupervised federated learning approach utilizing autoencoders, where each 5G core node independently trains a local model and a central server aggregates these models to improve detection performance. Similarly, SENTINEL~\cite{sentinel2023} presented a self-protecting control plane framework that leverages AI to identify and isolate malicious users responsible for launching DDoS attacks. DL techniques have also been explored, for example, Hussain et al.~\cite{hussain2020deep} developed a CNN-based system for detecting compromised network cells under attack, while the authors in~\cite{deepsecure2022} proposed a framework using LSTM networks to identify DDoS incidents in UE-related network traffic. Recent work has further applied sequence modeling for improved threat detection. Pell et al.~\cite{pell2024lstm} used LSTM networks to detect PFCP signaling attacks in 5G networks with high accuracy. Similarly, Djaidja et al.~\cite{Djaidja2024early} proposed attention-based RNNs for early intrusion detection by leveraging packet sequence data, reducing detection latency.

% \cite{ansam_fl_ids_survey} & 2024 & \checkmark & \xmark  & \xmark  & Survey of FL architectures, aggregation strategies, privacy, and challenges for IDS.  \\
%\hline
\renewcommand{\arraystretch}{1.05} 
\begin{table*}[!t]
\centering
\caption{Overview of Survey Papers on IDS, XAI, and LLM Integration}
\label{tab:survey_summary}
\begin{tabularx}{\textwidth}{|l|c|c|c|c|X|}
\hline
\textbf{Ref.} & \textbf{Year} & \textbf{ML/DL} &  \textbf{XAI} & \textbf{LLM} & \textbf{Focus} \\
\hline\hline
\cite{ahmad_study_ids} & 2021 & \checkmark & \xmark  & \xmark  & Review of ML/DL techniques for NIDS, coverage of datasets, evaluation metrics, and challenges. \\
\hline
\cite{Abdulganiyu_ids_survey} & 2023 & \checkmark & \xmark  & \xmark  & Classification of signature, anomaly, and hybrid IDS, emphasis on dataset and methodology usage. \\
\hline
\cite{xu2025deeplearningbasedintrusiondetection_survey}  & 2025 & \checkmark &  \xmark  & \xmark  & Review of DL-based IDS lifecycle, covering data management, attack
investigation, methodologies, and datasets.\\
\hline
\cite{he_advml_ids_survey} & 2023 & \checkmark &  \xmark  & \xmark  & Review of adversarial attacks and defenses for DL-based NIDS. \\
\hline
\cite{subashIDSxai}  & 2022 & \checkmark &  \checkmark  & \xmark  & Survey of XAI-IDSs, explainability of black/white-box models, and general X-IDS architecture.\\
\hline
\cite{alfredoXAIsurvey}  & 2024 & \checkmark &  \checkmark  & \xmark  & Review of XAI methods for network traffic classification and prediction. \\
\hline
\cite{moustafaXAIsurvey} & 2025 & \checkmark & \checkmark  & \xmark  & Survey of XAI methods for anomaly-based IDS in IoT,  \\
\hline
\cite{XAIsurveyB5G}  & 2022 & \checkmark &  \checkmark  & \checkmark  & Analysis of XAI solutions in 5G/B5G, highlighting challenges and future trends.   \\
\hline
\cite{hao2025LLMforTELCOSURVEY} & 2025 & \xmark & \xmark  & \checkmark  & Survey of LLM fundamentals, techniques, deployment strategies, and applications tailored to telecommunications.  \\
\hline
\cite{kheddar_llm_ids_survey} & 2025 & \xmark & \xmark  & \checkmark  & Review of Transformer and LLM-based IDS, applications and challenges.  \\
\hline
\cite{ferrag2024generative}  & 2024 & \xmark & \xmark  & \checkmark  & Review of LLMs in cybersecurity, including applications, vulnerabilities, mitigation strategies, and datasets.\\
\hline
\cite{xu2025largelanguagemodelscyber} & 2025 & \xmark & \xmark  & \checkmark  & Analysis of LLM applicability on various cybersecurity tasks. \\
\hline
\cite{djallel2024advancing}  & 2024 & \xmark & \checkmark  & \checkmark  & Review of LLM-based IDS methods, applications, challenges, and trends. \\
\hline
\end{tabularx}
\end{table*}
\renewcommand{\arraystretch}{1} 

\subsection{XAI for Intrusion Detection}
Although AI-driven IDSs have demonstrated notable improvements in detection accuracy, their opaque decision-making processes can impede the interpretability and validation of alerts by security analysts, potentially reducing trust and hampering incident response times. To address these limitations, recent survey studies~\cite{subashIDSxai},\cite{alfredoXAIsurvey}, \cite{moustafaXAIsurvey},\cite{XAIsurveyB5G} have explored the integration of XAI techniques with anomaly detection models in network security. These surveys collectively emphasize the growing interest in XAI approaches for enhancing interpretability, transparency, and decision support in AI-based IDSs.

Several recent studies have proposed explainable AI-based frameworks for anomaly detection in RANs. Barnard et al.~\cite{barnard2022xai} proposed a two-stage pipeline combining XGBoost-based supervised intrusion detection with SHAP explanations feeding a deep autoencoder for anomaly detection. Basaran and Dressler~\cite{basaran2025xainomaly} introduced XAInomaly, a semi-supervised Deep Contractive Autoencoder (DeepCAE) that integrates Generative AI and XAI to enhance scalability and interpretability while reducing the computational complexity of DL models. Similarly, Marantis et al.~\cite{marantisran} proposed the Explainable Anomaly Prediction (XAP) framework, which employs an LSTM Encoder-Decoder combined with an Isolation Forest and utilizes SHAP-based feature attribution for interpretable detection of RAN anomalies. In addition, Sun et al.~\cite{spotlight2024} developed SpotLight, a multi-stage explainable generative model aimed at detecting and localizing anomalies in large-scale Open RAN deployments. Additionally, Mahbooba et al.~\cite{mahbooba2021explainable} explored the use of decision tree models to enhance trust management in IDS through explainability, focusing on interpretable rules extracted from benchmark datasets. Gaspar et al.~\cite{gaspar2024xaiids} evaluated LIME and SHAP applicability on a multi-layer perceptron for IDS, demonstrating how these XAI methods improve interpretability and trustworthiness of black-box intrusion detection models. Furthermore, Wali et al.~\cite{wali2025xairf} proposed a reliable IDS combining explainable AI with random forest models to enhance resilience against adversarial attacks without retraining.

\subsection{LLM Agents for Intrusion Detection, Mitigation and Explainability}
Recently, the integration of LLMs and IDSs has attracted significant attention, with several comprehensive surveys highlighting their potential to enhance detection accuracy, scalability, and interpretability across a broad spectrum of cybersecurity applications~\cite{hao2025LLMforTELCOSURVEY}, \cite{kheddar_llm_ids_survey}, \cite{ferrag2024generative},\cite{xu2025largelanguagemodelscyber}, \cite{djallel2024advancing},\cite{harnessing_llms_ids_lira}. Table~\ref{tab:survey_summary} provides an overview of recent surveys covering ML/DL-based IDSs, XAI-IDS integration, and the use of LLMs for detection, mitigation and natural language insights.  Based on this foundation, recent research has demonstrated the effectiveness of LLM in code vulnerability detection~\cite{shestov2024finetuninglargelanguagemodels}  and intrusion detection, showing that models such as BERT can provide good performance in terms of accuracy, robustness, and even privacy preservation~\cite{LLM_based_IDS,ferrag2024revolutionizingcyberthreatdetection}. In the context of RANs, Moore et al.~\cite{moore2025integrated} proposed the use of LLMs as detection mechanisms, illustrating their applicability in intelligent network slicing and real-time security for O-RAN-enabled wireless deployments.

Beyond basic translation and summarization, recent advances in the reasoning capabilities of LLMs have unlocked new avenues for autonomous decision support in cybersecurity, including the automated proposal of mitigation strategies in response to detected threats~\cite{ferrag2025reasoninglimitsadvancesopen}. A notable research trend is the exploration of LLMs to enhance the explainability of AI-driven security systems, aiming to overcome the limitations of traditional XAI techniques, which often generate complex or overly technical outputs that are challenging for non-expert analysts to interpret. Recent studies have demonstrated a range of LLM-powered frameworks designed to address this gap. For instance, Wen et al.~\cite{6gxsec} introduced 6G-XSec, an explainable edge security framework that integrates LLMs with DL-based anomaly detection, improving security telemetry in O-RAN environments. Likewise, Mavrepisa et al.~\cite{mavrepis2024xaialllargelanguage} developed x-[plAIn], a GPT-based LLM that tailors XAI explanations for both technical users and non-experts, making model outputs clearer and more actionable. Similarly, Ali and Kostakos~\cite{ali2023huntgptintegratingmachinelearningbased} proposed HuntGPT, a dashboard that combines ML-based anomaly detection, XAI techniques (SHAP, LIME), and a GPT-3.5 Turbo conversational agent to translate complex threat detections into human-readable insights, improving decision-making and trust in AI-driven security systems. Furthermore, Liu et al.\cite{liu2024largelanguagemodelsdeliver} introduced an LLM-based framework that delivers accurate and interpretable anomaly detection along with comprehensive, human-readable explanations. Lastly, Jüttner et al.\cite{juttner2023chatidsexplainablecybersecurityusing} presented ChatIDS, which leverages LLMs to translate technical IDS alerts into intuitive language and actionable guidance specifically for non-expert users, thereby enhancing the accessibility and usability of cybersecurity analytics.

Although prior work has explored explainable anomaly detection or the use of LLMs for generating human-readable insights, they typically address these components separately. Table~\ref{tab:related_approaches} presents a comparison with recent related works. Our proposed approach uniquely integrates both through two dedicated applications: one for intrusion detection and another for post-hoc explainability and natural language interpretation. Although systems like HuntGPT employ a similar XAI–LLM pipeline, they are neither designed for nor deployable within Open RAN environments. Our approach fills this gap by enabling near-real-time detection while providing interpretable, human-readable insights through LLMs.

\begin{table*}[!t]
\centering
\caption{Comparison of IDS, XAI, and LLM Integration Across Related Work}
\label{tab:related_approaches}
\resizebox{\textwidth}{!}{%
\begin{tabular}{|l|c|c|c|c|c|c|c|c|c|}
\hline
\textbf{Feature} & \textbf{\cite{LLM_based_IDS}} & \textbf{\cite{ferrag2024revolutionizingcyberthreatdetection}} & \textbf{\cite{moore2025integrated}} & \textbf{\cite{6gxsec}} & \textbf{\cite{mavrepis2024xaialllargelanguage}} & \textbf{\cite{ali2023huntgptintegratingmachinelearningbased}} & \textbf{\cite{liu2024largelanguagemodelsdeliver}} & \textbf{\cite{juttner2023chatidsexplainablecybersecurityusing}} & \textbf{This Work} \\
\hline \hline
Detection Method          & LLM & LLM & LLM & AE/LSTM  & N/A & RF  & LLM & Signature based  & LSTM \\
XAI Integration           & \xmark  & \xmark & \xmark & \xmark & \checkmark & \checkmark  & \xmark & \xmark & \checkmark \\
LLM for Natural Language Insights & \xmark & \xmark & \xmark & \checkmark  & \checkmark & \checkmark  & \checkmark  &  \checkmark &  \checkmark \\
LLM for Mitigation Reasoning      & \xmark & \xmark & \xmark & \xmark & \xmark & \checkmark &  \xmark & \checkmark   &  \checkmark \\
Domain                    & General & IoT/IIoT  & Open RAN  & Open RAN  & General    & General  &  General  &  General  & Next-Gen RANs \\
\hline
\end{tabular}
}
\end{table*}

\section{Interpretable Anomaly-Based DDoS Detection System}\label{sec:3}
This section introduces our framework for intelligent DDoS detection and explainability within the Open RAN architecture. We integrate an LSTM-based detection model with XAI techniques and LLMs to provide interpretable insights.
Section~\ref{ssec:framework} details the architecture of the proposed framework and how it can be integrated into the Near-RT RIC. Section~\ref{ssec:xai} focuses on the local explanation methods used to understand individual predictions, while Section~\ref{ssec:llm} describes the LLM-based interpretability module.

\subsection{Proposed Framework}\label{ssec:framework}
\begin{figure}[!t]
    \centering
    \includegraphics[width=1.00\columnwidth]{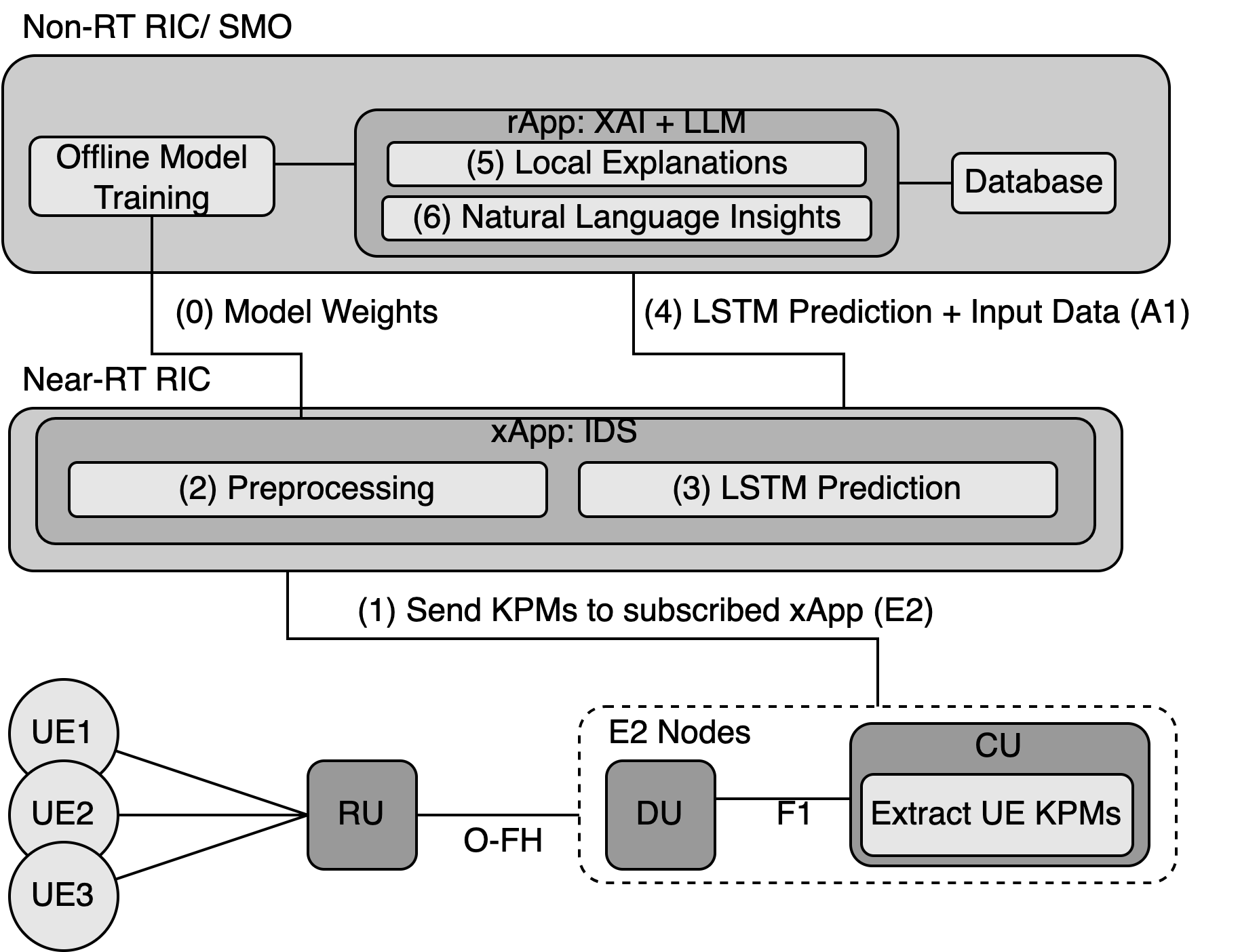}
    \caption{\small End-to-end workflow of the proposed framework within the Open RAN architecture}
    \label{fig:iris_pipeline}
\end{figure} 
The proposed framework is illustrated in Fig.~\ref{fig:iris_pipeline}. This framework integrates machine learning and explainability techniques within the Open RAN architecture to enable near real-time DDoS detection with interpretable insights. As traffic is exchanged from the UEs to the core network, the RAN components collect and extract KPMs for each UE. These KPMs are sent via the E2 interface (Step 1) to the Near-RT RIC, where they are received by the IDS xApp. The framework consists of one xApp and one rApp:

\begin{itemize}
    \item \textbf{xApp: IDS}, which includes a preprocessing module (Step 2) to organize KPMs into sequential input suitable for the LSTM model. The LSTM performs DDoS prediction (Step 3), and both the prediction output and the corresponding preprocessed instance are forwarded to the XAI–LLM rApp (via the A1 interface) for further interpretation.
    
    \item \textbf{rApp: XAI + LLM}, which receives the output of the IDS xApp (Step 4). It first applies local explanation methods (SHAP and LIME) to interpret the prediction (Step 5), and then uses an LLM to translate these explanations into human-readable insights (Step 6).
    The decision to implement this functionality as an rApp is driven by both computational and architectural factors. Large-scale LLMs (e.g., GPT, DeepSeek) introduce notable inference latency, which can exceed the near-real-time constraints of xApps. rApps, being suited for non-real-time operations, are better aligned with the compute-intensive nature of XAI+LLM processing. Moreover, since this module serves an advisory role, providing interpretability and mitigation suggestions without directly interfacing with RAN components, its design naturally aligns with the rApp framework. Future extensions that automate mitigation could reconsider an xApp implementation.

\end{itemize}
The final output can be saved in a database for future use. Additionally, the Non-RT RIC includes an offline model training module, which periodically updates the model weights and deploys them to the xApp and rApp (Step 0).

\subsection{XAI Methods}\label{ssec:xai}
XAI addresses the challenge of providing transparency in the decision-making process of machine learning models. These techniques can generally be grouped into two major categories: intrinsic and post-hoc methods. Intrinsic approaches involve the use of inherently explainable models, such as tree-based or linear models. However, these may struggle or be unsuitable when working with high-dimensional or time-series data. In contrast, post-hoc methods aim to explain already-trained black-box models, such as neural networks, by analyzing their behavior after training. Post-hoc explainability techniques can be further divided into model-specific and model-agnostic approaches. The former are tailored to a particular model type and often leverage internal structures, such as attention weights, to provide insights. The latter, such as SHAP (SHapley Additive exPlanations)~\cite{NIPS2017_8a20a862} or LIME (Local Interpretable Model-Agnostic Explanations)~\cite{ribeiro2016should}, treat the model as a black box and generate explanations without relying on its internal structure, making them applicable to a wide range of models. Finally, explanations can be categorized as local, which focus on individual predictions, or global, which aim to describe the overall behavior of the model.
\subsubsection{Local Explanations} To interpret individual predictions made by the LSTM model, two local explainability methods were examined, LIME and SHAP. \\ \textbf{LIME:} constructs a local surrogate model around a specific prediction to approximate the behavior of a complex black-box model. The idea is to perturb the input sample multiple times and observe how these changes affect the model's predictions. These perturbed samples are weighted by their similarity to the original instance, and a simple, interpretable model, typically a sparse linear model, is trained to mimic the local decision boundary. LIME formalizes this process as the following optimization problem
    \begin{equation}\small 
        \xi(x)  = \arg\min_{g \in G} \mathcal{L}(f, g, \pi_x) + \Omega(g),
    \end{equation}
    where \( f \) is the original black-box model, \( g \in G \) is the family of interpretable models, \( \mathcal{L}(f, g, \pi_x) \) measures how well \( g \) approximates \( f \) in the locality defined by \( \pi_x \), and \( \Omega(g) \) enforces simplicity in \( g \) (e.g., limiting the number of features). The solution \( \xi(x) \) corresponds to the explanation for instance \( x \). LIME is model-agnostic and flexible, making it useful for understanding individual predictions across a wide variety of models and data types~\cite{ribeiro2016should}. \\ \textbf{SHAP:} values are grounded in cooperative game theory and aim to fairly distribute a model’s output among its input features. Each feature is treated as a "player" in a game, and its SHAP value reflects the average contribution it makes to the prediction across all possible feature subsets. The SHAP value for a feature \( i \) is defined as
    \begin{equation}\small 
    \begin{split}
    \phi_i(f, x) = \sum_{S \subseteq F \setminus \{i\}} 
    \frac{|S|!(|F| - |S| - 1)!}{|F|!} \cdot \\
    \left[ f_{S \cup \{i\}}(x) - f_S(x) \right],
    \end{split}
    \end{equation}
    where \( F \) is the full set of input features, \( f_S(x) \) is the model’s output when only the subset \( S \) is present, and \( f_{S \cup \{i\}}(x) \) is the output when the feature \( i \) is added to the subset, representing the marginal contribution of feature \( i \). SHAP values offer both local and global explainability and can be used with any model, although exact computation can be expensive for large feature sets. Computing exact Shapley values requires evaluating all \( 2^n \) feature subsets, which becomes unmanageable for high-dimensional inputs such as time-series data. To overcome this, Kernel SHAP was used, an efficient, model-agnostic approximation that fits a weighted linear model on randomly sampled subsets while preserving the core properties of Shapley values.

\subsubsection{Global Explanations}
In our paper, we used the global shapley values to provide a global explanation of our model. The global importance of a feature \( i \) can be estimated by aggregating its SHAP values across all samples in the dataset. This is typically done by taking the mean of the absolute SHAP values

\begin{equation}\small 
    \Phi_i = \frac{1}{N} \sum_{j=1}^{N} |\phi_i^{(j)}|,
\end{equation}
where \( \phi_i^{(j)} \) is the SHAP value of feature \( i \) for the \( j \)-th instance, and \( N \) is the total number of instances. The resulting value \( \Phi_i \) reflects the average contribution (regardless of direction) of feature \( i \) to the model's predictions, offering a measure of global importance.

\subsection{LLM for Human Interpretability}\label{ssec:llm}
Outputs from XAI methods provide detailed insights into model decisions at the feature level, but they are often difficult to interpret or act upon, particularly for non-technical users. To bridge this gap, we introduce an LLM-based interpretability module designed to translate complex explanations into clear insights.

To enable consistent and context-aware interpretations, the input prompt provided to the LLM was carefully designed to encapsulate both global and instance-specific information, as detailed below. An example of such a prompt is shown in 
Fig.~\ref{fig:llm_input_zs} for reference. 
\begin{itemize}
    \item \textbf{General Feature Statistics:} A table showing the mean and standard deviation of each feature under both normal and attack conditions, providing global context about the data distribution.

    \item \textbf{LSTM Input Sequence:} A matrix representing the input to the model, consisting of 3 time steps and 14 features, highlighting the temporal behavior of the instance being evaluated.

    \item \textbf{Model Output:} The binary classification result (normal or anomalous) produced by the LSTM model for the given input sequence.

    \item \textbf{Local Explanation Tables:} Tabulated outputs from LIME and SHAP explanation methods, showing the contribution of individual features to the specific prediction made by the model.

    \item \textbf{Global Feature Importance Table:} A table of mean absolute SHAP values computed across the dataset and structured by feature and time step, enabling the LLM to understand which features are generally most influential.

     \item \textbf{Task Instructions:} A final instruction block asking the LLM to produce a human-readable summary of the model's explanation and to suggest potential mitigation strategies. For LLMs that support reasoning, we explore whether enabling such modes improves performance on this task.
\end{itemize}
Initially, we adopted a zero-shot prompting strategy, in which the LLM was presented with the structured input prompt and asked to generate explanations without any prior examples. This approach enabled rapid evaluation of interpretability without the need for task-specific fine-tuning.

To further improve the quality of the generated insights, we applied a few-shot prompting strategy. Such strategies can provide the model with the necessary contextual cues through examples, while requiring relatively small datasets and low computational resources~\cite{hao2025LLMPrompts, wang2025LLMSecurityPrivacy}. In this approach, we appended the prompt with a couple example input-output pairs, each consisting of the previous prompt input followed by a high-quality, human-written explanation. The inclusion of these demonstrations can help the LLM better understand the structure, tone, and content expected in its responses. Fig.~\ref{fig:llm_input_fs} provides the complete set of example inputs and corresponding human-written outputs used during the few-shot prompting process.

\onecolumn
\begin{tcolorbox}[colback=white, colframe=black, fonttitle=\bfseries, sharp corners, boxrule=0.4pt]
\textbf{System Prompt:} You are a cybersecurity expert analyzing LSTM outputs for DDoS detection. \\
\textbf{User Prompt:}\\
\scriptsize
\textbf{Normalized General Data Distribution (Feature Statistics):}
\begin{center}
\begin{tabular}{|l|c|c|c|c|}
\hline
\textbf{Feature} & \textbf{Normal Mean} & \textbf{Normal Std} & \textbf{Attack Mean} & \textbf{Attack Std} \\
\hline
epre & 0.5695 & 0.2032 & 0.4760 & 0.1560 \\\hline
pusch\_snr & 0.6561 & 0.0997 & 0.6250 & 0.0732 \\\hline
p\_ue & 0.3921 & 0.2464 & 0.3714 & 0.1638 \\\hline
ul\_mcs & 0.6775 & 0.2775 & 0.5885 & 0.2767 \\\hline
cqi & 0.8097 & 0.2172 & 0.6855 & 0.2649 \\\hline
ul\_bitrate & 0.0766 & 0.1215 & 0.0905 & 0.0983 \\\hline
dl\_mcs & 0.7158 & 0.2803 & 0.5847 & 0.3470 \\\hline
dl\_retx & 0.0455 & 0.0745 & 0.0541 & 0.0636 \\\hline
ul\_tx & 0.2864 & 0.3242 & 0.3674 & 0.1803 \\\hline
dl\_tx & 0.1623 & 0.2190 & 0.0790 & 0.0434 \\\hline
ul\_retx & 0.0864 & 0.1140 & 0.1630 & 0.1318 \\\hline
dl\_bitrate & 0.0162 & 0.0220 & 0.0006 & 0.0005 \\\hline
dl\_err & 0.0012 & 0.0085 & 0.0075 & 0.0262 \\\hline
ul\_err & 0.0008 & 0.0054 & 0.0233 & 0.0865 \\
\hline
\end{tabular}
\end{center}

\textbf{Input Sequence to LSTM Model (3 timesteps × 14 features):}\vspace{-1.5em}

\begin{center}
\resizebox{\textwidth}{!}{%
\begin{tabular}{|l|c| c| c| c| c| c| c| c| c| c| c| c| c| c|}
\hline
\textbf{Timestep} & epre & pusch\_snr & p\_ue & ul\_mcs & cqi & ul\_bitrate & dl\_mcs & dl\_retx & ul\_tx & dl\_tx & ul\_retx & dl\_bitrate & dl\_err & ul\_err \\
\hline
T0 & 0.32281 & 0.37859 & 0.45652 & 0.67315 & 1.00000 & 0.42860 & 0.995 & 0.00779 & 0.45002 & 0.15981 & 0.340 & 0.00276 & 0 & 0 \\\hline
T1 & 0.32105 & 0.36902 & 0.45652 & 0.66537 & 1.00000 & 0.42730 & 0.990 & 0.00390 & 0.45485 & 0.15981 & 0.348 & 0.00275 & 0 & 0 \\  \hline
T2 & 0.32105 & 0.37094 & 0.47826 & 0.66926 & 0.93333 & 0.42821 & 0.960 & 0.00260 & 0.45292 & 0.15910 & 0.340 & 0.00267 & 0 & 0 \\
\hline
\end{tabular}
}
\end{center}

\textbf{Model Prediction Output:} \texttt{1 (Anomalous)}

\textbf{Local Explanation Table (LIME):} 

\begin{center}
\begin{tabular}{|l r|l r|l r|}
\hline
\textbf{Feature} & \textbf{Contrib.} & \textbf{Feature} & \textbf{Contrib.} & \textbf{Feature} & \textbf{Contrib.} \\
\hline
ul\_bitrate\_t-0 \(>\) 0.22 & 0.09664 & ul\_retx\_t-0 \(>\) 0.30 & 0.07914 & ul\_bitrate\_t-2 \(>\) 0.22 & -0.07585 \\\hline
0.72 \(<\) dl\_mcs\_t-1 \(\leq\) 0.99 & -0.06507 & 0.01 \(<\) dl\_tx\_t-0 \(\leq\) 0.42 & 0.05576 & ul\_retx\_t-1 \(>\) 0.30 & 0.05342 \\\hline
0.14 \(<\) ul\_tx\_t-0 \(\leq\) 0.62 & 0.04732 & epre\_t-0 \(\leq\) 0.37 & 0.04471 & 0.01 \(<\) dl\_tx\_t-1 \(\leq\) 0.42 & 0.03819 \\\hline
ul\_retx\_t-2 \(>\) 0.30 & 0.03615 & 0.00 \(<\) dl\_retx\_t-2 \(\leq\) 0.01 & 0.03157 & 0.14 \(<\) ul\_tx\_t-1 \(\leq\) 0.62 & 0.02893 \\\hline
dl\_err\_t-1 \(\leq\) 0.00 & -0.02838 & epre\_t-2 \(\leq\) 0.36 & 0.02500 & ul\_bitrate\_t-1 \(>\) 0.22 & -0.02350 \\\hline
dl\_mcs\_t-2 \(>\) 0.99 & -0.02321 & 0.14 \(<\) ul\_tx\_t-2 \(\leq\) 0.62 & 0.02317 & 0.01 \(<\) dl\_tx\_t-2 \(\leq\) 0.42 & -0.02195 \\\hline
ul\_err\_t-0 \(\leq\) 0.00 & -0.02139 & 0.00 \(<\) dl\_bitrate\_t-0 \(\leq\) 0.06 & 0.01564 & 0.31 \(<\) p\_ue\_t-2 \(\leq\) 0.57 & -0.01205 \\\hline
epre\_t-1 \(\leq\) 0.36 & 0.01125 & dl\_retx\_t-1 \(\leq\) 0.00 & -0.01082 & 0.31 \(<\) p\_ue\_t-1 \(\leq\) 0.57 & -0.01081 \\\hline
pusch\_snr\_t-1 \(\leq\) 0.43 & -0.01075 & 0.31 \(<\) p\_ue\_t-0 \(\leq\) 0.57 & 0.00983 & dl\_err\_t-2 \(\leq\) 0.00 & 0.00944 \\\hline
dl\_retx\_t-0 \(\leq\) 0.00 & -0.00879 & ul\_err\_t-1 \(\leq\) 0.00 & -0.00749 & 0.00 \(<\) dl\_bitrate\_t-2 \(\leq\) 0.06 & 0.00664 \\\hline
0.87 \(<\) cqi\_t-0 \(\leq\) 1.00 & -0.00591 & 0.45 \(<\) ul\_mcs\_t-0 \(\leq\) 0.69 & -0.00561 & 0.00 \(<\) dl\_bitrate\_t-1 \(\leq\) 0.06 & 0.00521 \\\hline
pusch\_snr\_t-0 \(\leq\) 0.43 & -0.00492 & 0.45 \(<\) ul\_mcs\_t-1 \(\leq\) 0.69 & 0.00448 & 0.87 \(<\) cqi\_t-1 \(\leq\) 1.00 & -0.00388 \\\hline
0.45 \(<\) ul\_mcs\_t-2 \(\leq\) 0.69 & 0.00376 & dl\_err\_t-0 \(\leq\) 0.00 & -0.00259 & 0.72 \(<\) dl\_mcs\_t-0 \(\leq\) 0.99 & 0.00222 \\\hline
ul\_err\_t-2 \(\leq\) 0.00 & -0.00139 & 0.87 \(<\) cqi\_t-2 \(\leq\) 1.00 & 0.00084 & pusch\_snr\_t-2 \(\leq\) 0.43 & -0.00076 \\
\hline
\end{tabular}
\end{center}

\textbf{SHAP Local Heatmap:}\vspace{-1.5em}
\begin{center}
\resizebox{\textwidth}{!}{%
\begin{tabular}{|l|c| c| c| c| c| c| c| c| c| c| c |c| c| c|}
\hline
\textbf{Timestep} & epre & pusch\_snr & p\_ue & ul\_mcs & cqi & ul\_bitrate & dl\_mcs & dl\_retx & ul\_tx & dl\_tx & ul\_retx & dl\_bitrate & dl\_err & ul\_err \\
\hline
T0 & -0.0159 & -0.00667 & 0.00869 & 0.0105 & 0.00236 & -0.01269 & 0 & 0.02119 & 0.02953 & 0.07447 & 0.01947 & 0.08598 & 0 & -0.00429 \\\hline
T1 & 0 & -0.01671 & -0.01998 & 0.01434 & 0 & 0.03824 & -0.01903 & 0.01008 & 0.02105 & 0.14609 & 0.03622 & 0.06092 & 0 & 0.00252 \\ \hline
T2 & 0.01386 & 0 & -0.01772 & -0.01354 & 0.00369 & 0.1649 & -0.00311 & 0.00705 & 0.13354 & 0.15929 & 0.04019 & 0.02695 & 0 & -0.00749 \\
\hline
\end{tabular}
}
\end{center}

\textbf{Global SHAP Feature Importance:}\vspace{-0.8em}
\begin{center}
\resizebox{\textwidth}{!}{%
\begin{tabular}{|l|c|c|c| c| c |c| c |c| c| c| c| c| c |c|}
\hline
\textbf{Timestep} & epre & pusch\_snr & p\_ue & ul\_mcs & cqi & ul\_bitrate & dl\_mcs & dl\_retx & ul\_tx & dl\_tx & ul\_retx & dl\_bitrate & dl\_err & ul\_err \\
\hline
T0 & 0.00975 & 0.00844 & 0.00853 & 0.01271 & 0.00466 & 0.01740 & 0.01768 & 0.01208 & 0.01597 & 0.09860 & 0.01823 & 0.03002 & 0.00051 & 0.00514 \\\hline
T1 & 0.00799 & 0.00981 & 0.01421 & 0.00769 & 0.00560 & 0.01767 & 0.03328 & 0.01418 & 0.02568 & 0.07379 & 0.02140 & 0.01703 & 0.00058 & 0.00429 \\\hline
T2 & 0.00549 & 0.00538 & 0.00705 & 0.00464 & 0.00482 & 0.04328 & 0.01549 & 0.01309 & 0.09734 & 0.06278 & 0.01995 & 0.01134 & 0.00053 & 0.00468 \\
\hline
\end{tabular}
}
\end{center}

\textbf{Task for the LLM:} \\
Provide a human-readable summary of the model's decision. Highlight the most influential features or patterns that contributed to the classification. Assess the likelihood of misclassification and suggest actionable mitigation strategies for the network operator.
\end{tcolorbox}
\captionof{figure}{Zero-Shot LLM Input Prompt}
\label{fig:llm_input_zs}

\clearpage
\noindent
\begin{tcolorbox}[colback=white, colframe=black, fonttitle=\bfseries, sharp corners, boxrule=0.5pt, breakable , width=\textwidth]

\textbf{System Prompt:} You are a cybersecurity expert analyzing LSTM outputs for DDoS detection. Respond with 3 sections: Anomaly Summary (3-6 bullet points), Misclassification Likelihood (1-2 bullets), and Mitigation Steps (2-4 bullets).

\textbf{User Prompt:}
\scriptsize

\textbf{Normalized General Data Distribution (Feature Statistics):}

\begin{center}
\begin{tabular}{|l|c|c|c|c|}
\hline
\textbf{Feature} & \textbf{Normal Mean} & \textbf{Normal Std} & \textbf{Attack Mean} & \textbf{Attack Std} \\
\hline
epre & 0.5695 & 0.2032 & 0.4760 & 0.1560 \\\hline
pusch\_snr & 0.6561 & 0.0997 & 0.6250 & 0.0732 \\\hline
p\_ue & 0.3921 & 0.2464 & 0.3714 & 0.1638 \\\hline
ul\_mcs & 0.6775 & 0.2775 & 0.5885 & 0.2767 \\\hline
cqi & 0.8097 & 0.2172 & 0.6855 & 0.2649 \\\hline
ul\_bitrate & 0.0766 & 0.1215 & 0.0905 & 0.0983 \\\hline
dl\_mcs & 0.7158 & 0.2803 & 0.5847 & 0.3470 \\\hline
dl\_retx & 0.0455 & 0.0745 & 0.0541 & 0.0636 \\\hline
ul\_tx & 0.2864 & 0.3242 & 0.3674 & 0.1803 \\\hline
dl\_tx & 0.1623 & 0.2190 & 0.0790 & 0.0434 \\\hline
ul\_retx & 0.0864 & 0.1140 & 0.1630 & 0.1318 \\\hline
dl\_bitrate & 0.0162 & 0.0220 & 0.0006 & 0.0005 \\\hline
dl\_err & 0.0012 & 0.0085 & 0.0075 & 0.0262 \\\hline
ul\_err & 0.0008 & 0.0054 & 0.0233 & 0.0865 \\
\hline
\end{tabular}
\end{center}

\textbf{Input Sequence to LSTM Model (3 timesteps × 14 features):}\vspace{-1.5em}

\begin{center}
\resizebox{\textwidth}{!}{
\begin{tabular}{|l|c|c|c|c|c|c|c|c|c|c|c|c|c|c|}
\hline
Timestep & epre & pusch\_snr & p\_ue & ul\_mcs & cqi & ul\_bitrate & dl\_mcs & dl\_retx & ul\_tx & dl\_tx & ul\_retx & dl\_bitrate & dl\_err & ul\_err \\ \hline
T0 & 0.74737 & 0.57553 & 0.58696 & 1.0 & 1.0 & 0.60758 & 0.985 & 0.00779 & 0.98600 & 0.92063 & 0.008 & 0.09832 & 0 & 0 \\ \hline
T1 & 0.74912 & 0.57744 & 0.58696 & 1.0 & 1.0 & 0.60748 & 0.985 & 0.01169 & 0.98648 & 0.93636 & 0.012 & 0.09832 & 0 & 0 \\ \hline
T2 & 0.74035 & 0.54493 & 0.54348 & 1.0 & 1.0 & 0.60909 & 0.985 & 0.01299 & 0.98503 & 0.93851 & 0.004 & 0.09836 & 0 & 0 \\  \hline
\end{tabular}
}
\end{center}

\textbf{Model Prediction Output:} \texttt{0 (Normal)}

\textbf{LIME Local Explanation:}

\begin{center}
\begin{tabular}{|l r|l r|l r|}
\hline
\textbf{Feature} & \textbf{Contrib.} & \textbf{Feature} & \textbf{Contrib.} & \textbf{Feature} & \textbf{Contrib.} \\
\hline
dl\_tx\_t-2 \(> 0.42\) & -0.27307 & dl\_tx\_t-1 \(> 0.42\) & -0.18314 & ul\_tx\_t-0 \(> 0.62\) & 0.15454 \\
\hline
dl\_tx\_t-0 \(> 0.42\) & -0.10833 & ul\_tx\_t-1 \(> 0.62\) & 0.09584 & dl\_bitrate\_t-2 \(> 0.06\) & -0.09024 \\
\hline
ul\_bitrate\_t-2 \(> 0.22\) & -0.08377 & dl\_bitrate\_t-1 \(> 0.06\) & -0.07593 & \(0.72 <\) dl\_mcs\_t-1 \(\leq 0.99\) & -0.06363 \\
\hline
ul\_bitrate\_t-0 \(> 0.22\) & 0.06249 & dl\_bitrate\_t-0 \(> 0.06\) & -0.05761 & p\_ue\_t-1 \(> 0.57\) & -0.04190 \\
\hline
ul\_tx\_t-2 \(> 0.62\) & 0.03369 & ul\_bitrate\_t-1 \(> 0.22\) & -0.02462 & epre\_t-1 \(> 0.75\) & -0.02422 \\
\hline
\(0.00 <\) ul\_retx\_t-0 \(\leq 0.04\) & -0.02136 & \(0.00 <\) ul\_retx\_t-1 \(\leq 0.04\) & -0.02122 & \(0.00 <\) dl\_retx\_t-1 \(\leq 0.01\) & -0.01804 \\
\hline
ul\_mcs\_t-0 \(> 0.96\) & 0.01601 & \(0.00 <\) dl\_retx\_t-2 \(\leq 0.01\) & 0.01566 & ul\_mcs\_t-2 \(> 0.96\) & -0.01502 \\
\hline
\(0.72 <\) dl\_mcs\_t-2 \(\leq 0.99\) & -0.01500 & \(0.31 <\) p\_ue\_t-0 \(\leq 0.57\) & 0.01380 & dl\_err\_t-0 \(\leq 0.00\) & 0.01311 \\
\hline
dl\_err\_t-2 \(\leq 0.00\) & -0.01300 & ul\_err\_t-0 \(\leq 0.00\) & -0.01174 & \(0.55 <\) epre\_t-2 \(\leq 0.75\) & 0.01065 \\
\hline
\(0.72 <\) dl\_mcs\_t-0 \(\leq 0.99\) & -0.00995 & ul\_mcs\_t-1 \(> 0.96\) & -0.00959 & \(0.52 <\) pusch\_snr\_t-1 \(\leq 0.65\) & -0.00901 \\
\hline
p\_ue\_t-2 \(> 0.57\) & 0.00869 & \(0.00 <\) ul\_retx\_t-2 \(\leq 0.04\) & -0.00780 & dl\_err\_t-1 \(\leq 0.00\) & -0.00582 \\
\hline
ul\_err\_t-2 \(\leq 0.00\) & -0.00553 & \(0.00 <\) dl\_retx\_t-0 \(\leq 0.01\) & -0.00360 & \(0.87 <\) cqi\_t-2 \(\leq 1.00\) & -0.00293 \\
\hline
\(0.52 <\) pusch\_snr\_t-0 \(\leq 0.65\) & 0.00237 & \(0.87 <\) cqi\_t-0 \(\leq 1.00\) & -0.00123 & \(0.52 <\) pusch\_snr\_t-2 \(\leq 0.65\) & -0.00088 \\
\hline
ul\_err\_t-1 \(\leq 0.00\) & -0.00082 & \(0.55 <\) epre\_t-0 \(\leq 0.75\) & -0.00080 & \(0.87 <\) cqi\_t-1 \(\leq 1.00\) & 0.00036 \\
\hline
\end{tabular}
\end{center}

\textbf{SHAP Local Heatmap:}\vspace{-0.75em}
\begin{center}
\resizebox{\textwidth}{!}{
\begin{tabular}{|c|c|c|c|c|c|c|c|c|c|c|c|c|c|c|}
\hline
\textbf{Timestamp} & \textbf{epre} & \textbf{pusch\_snr} & \textbf{p\_ue} & \textbf{ul\_mcs} & \textbf{cqi} & \textbf{ul\_bitrate} & \textbf{dl\_mcs} & \textbf{dl\_retx} & \textbf{ul\_tx} & \textbf{dl\_tx} & \textbf{ul\_retx} & \textbf{dl\_bitrate} & \textbf{dl\_err} & \textbf{ul\_err} \\
\hline
T0 & 0.00738 & 0.0 & -0.00318 & 0.0 & 0.00435 & 0.0 & 0.0 & 0.0 & 0.00554 & -0.05466 & 0.00329 & 0.0 & 0.0 & 0.0 \\
\hline
T1 & -0.0084 & -0.00241 & -0.00809 & -0.00414 & 0.0 & 0.01005 & 0.0 & 0.0 & 0.02725 & -0.0193 & -0.00505 & 0.0 & 0.0 & -0.0034 \\
\hline
T2 & 0.0 & 0.0 & -0.0073 & -0.00656 & 0.00989 & 0.04504 & -0.00854 & 0.0 & 0.06172 & -0.03838 & 0.0 & -0.00638 & 0.0 & 0.0 \\
\hline
\end{tabular}
}
\end{center}

\textbf{Global SHAP Feature Importance:}\vspace{-0.75em}
\begin{center}
\resizebox{\textwidth}{!}{
\begin{tabular}{|c|c|c|c|c|c|c|c|c|c|c|c|c|c|c|}
\hline
\textbf{Timestamp} & \textbf{epre} & \textbf{pusch\_snr} & \textbf{p\_ue} & \textbf{ul\_mcs} & \textbf{cqi} & \textbf{ul\_bitrate} & \textbf{dl\_mcs} & \textbf{dl\_retx} & \textbf{ul\_tx} & \textbf{dl\_tx} & \textbf{ul\_retx} & \textbf{dl\_bitrate} & \textbf{dl\_err} & \textbf{ul\_err} \\
\hline
T0 & 0.00975 & 0.00844 & 0.00853 & 0.01271 & 0.00466 & 0.0174 & 0.01768 & 0.01208 & 0.01597 & 0.0986 & 0.01823 & 0.03002 & 0.00051 & 0.00514 \\
\hline
T1 & 0.00799 & 0.00981 & 0.01421 & 0.00769 & 0.0056 & 0.01767 & 0.03328 & 0.01418 & 0.02568 & 0.07379 & 0.0214 & 0.01703 & 0.00058 & 0.00429 \\
\hline
T2 & 0.00549 & 0.00538 & 0.00705 & 0.00464 & 0.00482 & 0.04328 & 0.01549 & 0.01309 & 0.09734 & 0.06278 & 0.01995 & 0.01134 & 0.00053 & 0.00468 \\
\hline
\end{tabular}
}
\end{center}

\textbf{Task For LLM:} \\
Provide a human-readable summary of the model's decision. Highlight which features or patterns most influenced the classification outcome. Assess the likelihood of misclassification and suggest actionable mitigation steps for the network operator.

\normalsize
\textbf{Assistant Response:}

\scriptsize
\textbf{Anomaly Summary:} The model classified this sequence as \textbf{normal}, reflecting balanced and efficient network behavior over three timesteps of normalized input features processed by an LSTM model.

\begin{itemize}
    \item Uplink and downlink activity are both high and symmetrical. \texttt{ul\_tx} and \texttt{dl\_tx} values range from $\sim$0.92 to 0.98, well above normal and attack means, but balanced across directions. This symmetry is reinforced by LIME, where \texttt{dl\_tx\_t-2}, \texttt{dl\_tx\_t-1}, and \texttt{dl\_tx\_t-0} contribute strongly to the normal prediction (-0.2731, -0.1831, -0.1083).
    \item Throughput metrics are elevated but not indicative of attack. \texttt{ul\_bitrate} values ($\sim$0.61) are above normal (0.08) and attack (0.09) means but are paired with high \texttt{dl\_bitrate} ($\sim$0.098), which is over 6× the normal mean and far above the attack average (0.0006). LIME assigns negative contributions to all \texttt{dl\_bitrate} timesteps (e.g., -0.0902 at T2), supporting the normal classification.
    \item Retransmissions are low and stable. \texttt{ul\_retx} and \texttt{dl\_retx} stay between 0.004 and 0.012, well below typical attack means (0.163 and 0.0541). LIME shows these features decreasing anomaly risk (e.g., \texttt{ul\_retx\_t-0}: -0.0214).
    \item Signal quality is strong. \texttt{epre} values ($\sim$0.74) exceed both normal (0.57) and attack (0.48) means. \texttt{pusch\_snr} remains above 0.54, and \texttt{cqi} is 1.0 across all steps. SHAP shows these features either neutral or mildly supportive of normal behavior.
    \item Error rates are zero (\texttt{ul\_err}, \texttt{dl\_err} = 0), ruling out radio link instability.
    \item SHAP local values confirm strong negative contributions from \texttt{dl\_tx} at every timestep (e.g., -0.0547 at T0, -0.0384 at T2), and modest positive impact from \texttt{ul\_tx} (e.g., +0.0617 at T2), preserving the balance. Global SHAP importance supports this interpretation—the most influential features (\texttt{ul\_tx}, \texttt{dl\_tx}, \texttt{dl\_bitrate}) behave in line with known normal patterns.
\end{itemize}

\textbf{Misclassification Likelihood:} \\
\textit{Very low.} All high-throughput features show symmetry and efficiency, with no signs of retransmission spikes or traffic asymmetry. Signal conditions are excellent, and LIME/SHAP contributions consistently align with normal behavior.

\textbf{Recommended Mitigation Steps:}
\begin{itemize}
    \item No action needed. This flow reflects well-functioning, high-throughput activity.
    \item Optionally log as a reference case for tuning thresholds or validating model reliability.
    \item Continue monitoring for changes over time or shifts in traffic symmetry.
\end{itemize}

\normalsize
\vspace{0.5em}
\textbf{User Prompt:}
\scriptsize

\textbf{Normalized General Data Distribution (Feature Statistics):}

\begin{center}
\begin{tabular}{|l|c|c|c|c|}
\hline
\textbf{Feature} & \textbf{Normal Mean} & \textbf{Normal Std} & \textbf{Attack Mean} & \textbf{Attack Std} \\
\hline
epre & 0.5695 & 0.2032 & 0.4760 & 0.1560 \\\hline
pusch\_snr & 0.6561 & 0.0997 & 0.6250 & 0.0732 \\\hline
p\_ue & 0.3921 & 0.2464 & 0.3714 & 0.1638 \\\hline
ul\_mcs & 0.6775 & 0.2775 & 0.5885 & 0.2767 \\\hline
cqi & 0.8097 & 0.2172 & 0.6855 & 0.2649 \\\hline
ul\_bitrate & 0.0766 & 0.1215 & 0.0905 & 0.0983 \\\hline
dl\_mcs & 0.7158 & 0.2803 & 0.5847 & 0.3470 \\\hline
dl\_retx & 0.0455 & 0.0745 & 0.0541 & 0.0636 \\\hline
ul\_tx & 0.2864 & 0.3242 & 0.3674 & 0.1803 \\\hline
dl\_tx & 0.1623 & 0.2190 & 0.0790 & 0.0434 \\\hline
ul\_retx & 0.0864 & 0.1140 & 0.1630 & 0.1318 \\\hline
dl\_bitrate & 0.0162 & 0.0220 & 0.0006 & 0.0005 \\\hline
dl\_err & 0.0012 & 0.0085 & 0.0075 & 0.0262 \\\hline
ul\_err & 0.0008 & 0.0054 & 0.0233 & 0.0865 \\
\hline
\end{tabular}
\end{center}

\textbf{Input Sequence to LSTM Model (3 timesteps × 14 features):}\vspace{-1.5em}

\begin{center}
\resizebox{\textwidth}{!}{%
\begin{tabular}{|l|c| c| c| c| c| c |c |c| c| c| c| c |c| c|}
\hline
\textbf{Timestep} & epre & pusch\_snr & p\_ue & ul\_mcs & cqi & ul\_bitrate & dl\_mcs & dl\_retx & ul\_tx & dl\_tx & ul\_retx & dl\_bitrate & dl\_err & ul\_err \\
\hline
T0 & 0.32281 & 0.37859 & 0.45652 & 0.67315 & 1.00000 & 0.42860 & 0.995 & 0.00779 & 0.45002 & 0.15981 & 0.340 & 0.00276 & 0 & 0 \\\hline
T1 & 0.32105 & 0.36902 & 0.45652 & 0.66537 & 1.00000 & 0.42730 & 0.990 & 0.00390 & 0.45485 & 0.15981 & 0.348 & 0.00275 & 0 & 0 \\\hline
T2 & 0.32105 & 0.37094 & 0.47826 & 0.66926 & 0.93333 & 0.42821 & 0.960 & 0.00260 & 0.45292 & 0.15910 & 0.340 & 0.00267 & 0 & 0 \\
\hline
\end{tabular}
}
\end{center}

\textbf{Model Prediction Output:} \texttt{1 (Anomalous)}

\textbf{LIME Local Explanation:}

\begin{center}
\begin{tabular}{|l r|l r|l r|}
\hline
\textbf{Feature} & \textbf{Contrib.} & \textbf{Feature} & \textbf{Contrib.} & \textbf{Feature} & \textbf{Contrib.} \\
\hline
ul\_bitrate\_t-0 \(>\) 0.22 & 0.09664 & ul\_retx\_t-0 \(>\) 0.30 & 0.07914 & ul\_bitrate\_t-2 \(>\) 0.22 & -0.07585 \\\hline
0.72 \(<\) dl\_mcs\_t-1 \(\leq\) 0.99 & -0.06507 & 0.01 \(<\) dl\_tx\_t-0 \(\leq\) 0.42 & 0.05576 & ul\_retx\_t-1 \(>\) 0.30 & 0.05342 \\\hline
0.14 \(<\) ul\_tx\_t-0 \(\leq\) 0.62 & 0.04732 & epre\_t-0 \(\leq\) 0.37 & 0.04471 & 0.01 \(<\) dl\_tx\_t-1 \(\leq\) 0.42 & 0.03819 \\\hline
ul\_retx\_t-2 \(>\) 0.30 & 0.03615 & 0.00 \(<\) dl\_retx\_t-2 \(\leq\) 0.01 & 0.03157 & 0.14 \(<\) ul\_tx\_t-1 \(\leq\) 0.62 & 0.02893 \\\hline
dl\_err\_t-1 \(\leq\) 0.00 & -0.02838 & epre\_t-2 \(\leq\) 0.36 & 0.02500 & ul\_bitrate\_t-1 \(>\) 0.22 & -0.02350 \\\hline
dl\_mcs\_t-2 \(>\) 0.99 & -0.02321 & 0.14 \(<\) ul\_tx\_t-2 \(\leq\) 0.62 & 0.02317 & 0.01 \(<\) dl\_tx\_t-2 \(\leq\) 0.42 & -0.02195 \\\hline
ul\_err\_t-0 \(\leq\) 0.00 & -0.02139 & 0.00 \(<\) dl\_bitrate\_t-0 \(\leq\) 0.06 & 0.01564 & 0.31 \(<\) p\_ue\_t-2 \(\leq\) 0.57 & -0.01205 \\\hline
epre\_t-1 \(\leq\) 0.36 & 0.01125 & dl\_retx\_t-1 \(\leq\) 0.00 & -0.01082 & 0.31 \(<\) p\_ue\_t-1 \(\leq\) 0.57 & -0.01081 \\\hline
pusch\_snr\_t-1 \(\leq\) 0.43 & -0.01075 & 0.31 \(<\) p\_ue\_t-0 \(\leq\) 0.57 & 0.00983 & dl\_err\_t-2 \(\leq\) 0.00 & 0.00944 \\\hline
dl\_retx\_t-0 \(\leq\) 0.00 & -0.00879 & ul\_err\_t-1 \(\leq\) 0.00 & -0.00749 & 0.00 \(<\) dl\_bitrate\_t-2 \(\leq\) 0.06 & 0.00664 \\\hline
0.87 \(<\) cqi\_t-0 \(\leq\) 1.00 & -0.00591 & 0.45 \(<\) ul\_mcs\_t-0 \(\leq\) 0.69 & -0.00561 & 0.00 \(<\) dl\_bitrate\_t-1 \(\leq\) 0.06 & 0.00521 \\\hline
pusch\_snr\_t-0 \(\leq\) 0.43 & -0.00492 & 0.45 \(<\) ul\_mcs\_t-1 \(\leq\) 0.69 & 0.00448 & 0.87 \(<\) cqi\_t-1 \(\leq\) 1.00 & -0.00388 \\\hline
0.45 \(<\) ul\_mcs\_t-2 \(\leq\) 0.69 & 0.00376 & dl\_err\_t-0 \(\leq\) 0.00 & -0.00259 & 0.72 \(<\) dl\_mcs\_t-0 \(\leq\) 0.99 & 0.00222 \\\hline
ul\_err\_t-2 \(\leq\) 0.00 & -0.00139 & 0.87 \(<\) cqi\_t-2 \(\leq\) 1.00 & 0.00084 & pusch\_snr\_t-2 \(\leq\) 0.43 & -0.00076 \\
\hline
\end{tabular}
\end{center}

\textbf{SHAP Local Heatmap:}\vspace{-1.5em}
\begin{center}
\resizebox{\textwidth}{!}{%
\begin{tabular}{|l|c| c| c |c| c| c| c |c| c| c| c| c| c| c|}
\hline
\textbf{Timestamp} & epre & pusch\_snr & p\_ue & ul\_mcs & cqi & ul\_bitrate & dl\_mcs & dl\_retx & ul\_tx & dl\_tx & ul\_retx & dl\_bitrate & dl\_err & ul\_err \\
\hline
T0 & -0.0159 & -0.00667 & 0.00869 & 0.0105 & 0.00236 & -0.01269 & 0 & 0.02119 & 0.02953 & 0.07447 & 0.01947 & 0.08598 & 0 & -0.00429 \\ \hline
T1 & 0 & -0.01671 & -0.01998 & 0.01434 & 0 & 0.03824 & -0.01903 & 0.01008 & 0.02105 & 0.14609 & 0.03622 & 0.06092 & 0 & 0.00252 \\\hline
T2 & 0.01386 & 0 & -0.01772 & -0.01354 & 0.00369 & 0.16490 & -0.00311 & 0.00705 & 0.13354 & 0.15929 & 0.04019 & 0.02695 & 0 & -0.00749 \\
\hline
\end{tabular}
}
\end{center}

\textbf{Global SHAP Feature Importance:}\vspace{-1.5em}
\begin{center}
\resizebox{\textwidth}{!}{%
\begin{tabular}{|l|c| c| c| c| c |c| c| c| c| c |c| c| c |c|}
\hline
\textbf{Timestamp} & epre & pusch\_snr & p\_ue & ul\_mcs & cqi & ul\_bitrate & dl\_mcs & dl\_retx & ul\_tx & dl\_tx & ul\_retx & dl\_bitrate & dl\_err & ul\_err \\
\hline
T0 & 0.00975 & 0.00844 & 0.00853 & 0.01271 & 0.00466 & 0.01740 & 0.01768 & 0.01208 & 0.01597 & 0.09860 & 0.01823 & 0.03002 & 0.00051 & 0.00514 \\  \hline
T1 & 0.00799 & 0.00981 & 0.01421 & 0.00769 & 0.00560 & 0.01767 & 0.03328 & 0.01418 & 0.02568 & 0.07379 & 0.02140 & 0.01703 & 0.00058 & 0.00429 \\\hline
T2 & 0.00549 & 0.00538 & 0.00705 & 0.00464 & 0.00482 & 0.04328 & 0.01549 & 0.01309 & 0.09734 & 0.06278 & 0.01995 & 0.01134 & 0.00053 & 0.00468 \\
\hline
\end{tabular}
}
\end{center}

\textbf{Task For LLM:} \\
Provide a human-readable summary of the model's decision. Highlight which features or patterns most influenced the classification outcome. Assess the likelihood of misclassification and suggest actionable mitigation steps for the network operator.

\normalsize
\textbf{Assistant Response:}

\scriptsize
\begin{itemize}
    \item \textbf{Anomaly Summary:} The model classified this sequence as \textbf{anomalous}, largely due to persistent high uplink activity and low downlink usage—a typical pattern in DDoS traffic. The analysis is based on three timesteps of normalized network features processed by an LSTM sequence model.
    \begin{itemize}
        \item All timesteps show elevated \texttt{ul\_bitrate} values around 0.43—over 3.5× the normal mean (0.08), and above the attack mean (0.09).
        \item \texttt{ul\_retx} values are consistently high ($\sim$0.34), over double the normal mean (0.086), matching the attack mean (0.16).
        \item \textbf{LIME insights}:
        \begin{itemize}
            \item \texttt{ul\_bitrate\_t-0 > 0.22}: +0.0966
            \item \texttt{ul\_retx\_t-0 > 0.30}: +0.0791
            \item \texttt{ul\_tx\_t-0}, \texttt{dl\_tx\_t-0}: +0.0473, +0.0558
        \end{itemize}
        \item \textbf{SHAP confirmation}:
        \begin{itemize}
            \item \texttt{ul\_bitrate\_t2}: +0.1649, \texttt{ul\_tx\_t2}: +0.1335, \texttt{dl\_tx\_t2}: +0.1593
            \item \texttt{dl\_bitrate} remains very low ($\sim$0.0027), more aligned with the attack mean (0.0006) than the normal (0.0162).
            \item \texttt{dl\_tx} is stable at $\sim$0.16—non-zero but much lower than attack burstiness, creating uplink-downlink imbalance.
        \end{itemize}
        \item \textbf{Signal and Error Context}:
        \begin{itemize}
            \item \texttt{epre} ($\sim$0.32) is closer to the attack mean (0.48) than the normal (0.57).
            \item \texttt{cqi} at 1.0 and high \texttt{dl\_mcs} ($\sim$0.99) indicate strong link quality.
            \item SHAP shows small or negative impact: \texttt{epre\_t0}: -0.0159, \texttt{pusch\_snr\_t1}: -0.0167, \texttt{cqi\_t0}: -0.0059.
            \item \texttt{ul\_err} and \texttt{dl\_err} are zero; LIME shows mild negative influence, e.g., \texttt{ul\_err\_t-0}: -0.0214.
        \end{itemize}
        \item Global SHAP confirms top influence from \texttt{ul\_bitrate}, \texttt{ul\_tx}, \texttt{dl\_tx}, validating local prediction.
    \end{itemize}

    \item \textbf{Misclassification Likelihood:} Low. The consistent uplink-heavy profile, combined with elevated retransmissions and low downlink bitrate, aligns strongly with known DDoS patterns. While signal metrics are clean, benign explanations are unlikely.

    \item \textbf{Recommended Mitigation Steps:}
    \begin{itemize}
        \item Apply uplink rate limiting or QoS shaping to reduce bandwidth impact if traffic persists.
        \item Correlate with flow logs or packet captures to verify session purpose and remote endpoints.
        \item If behavior is confirmed benign, mark this instance for future model retraining or labeling correction.
        \item Monitor for propagation—if multiple users show similar signatures, this may indicate a distributed attack.
    \end{itemize}
\end{itemize}

\vspace{0.5em}
\normalsize
\textbf{Actual Query User Prompt:}

\scriptsize

\textbf{Normalized General Data Distribution (Feature Statistics):}

\begin{center}
\begin{tabular}{|l|c|c|c|c|}
\hline
\textbf{Feature} & \textbf{Normal Mean} & \textbf{Normal Std} & \textbf{Attack Mean} & \textbf{Attack Std} \\
\hline
epre & 0.5695 & 0.2032 & 0.4760 & 0.1560 \\\hline
pusch\_snr & 0.6561 & 0.0997 & 0.6250 & 0.0732 \\\hline
p\_ue & 0.3921 & 0.2464 & 0.3714 & 0.1638 \\\hline
ul\_mcs & 0.6775 & 0.2775 & 0.5885 & 0.2767 \\\hline
cqi & 0.8097 & 0.2172 & 0.6855 & 0.2649 \\\hline
ul\_bitrate & 0.0766 & 0.1215 & 0.0905 & 0.0983 \\\hline
dl\_mcs & 0.7158 & 0.2803 & 0.5847 & 0.3470 \\\hline
dl\_retx & 0.0455 & 0.0745 & 0.0541 & 0.0636 \\\hline
ul\_tx & 0.2864 & 0.3242 & 0.3674 & 0.1803 \\\hline
dl\_tx & 0.1623 & 0.2190 & 0.0790 & 0.0434 \\\hline
ul\_retx & 0.0864 & 0.1140 & 0.1630 & 0.1318 \\\hline
dl\_bitrate & 0.0162 & 0.0220 & 0.0006 & 0.0005 \\\hline
dl\_err & 0.0012 & 0.0085 & 0.0075 & 0.0262 \\\hline
ul\_err & 0.0008 & 0.0054 & 0.0233 & 0.0865 \\
\hline
\end{tabular}
\end{center}

\textbf{Input Sequence to LSTM Model (3 timesteps × 14 features):}\vspace{-0.8em}

\begin{center}
\resizebox{\textwidth}{!}{%
\begin{tabular}{|l|c| c |c| c| c| c| c| c| c| c| c |c| c| c|}
\hline
\textbf{Timestep} & epre & pusch\_snr & p\_ue & ul\_mcs & cqi & ul\_bitrate & dl\_mcs & dl\_retx & ul\_tx & dl\_tx & ul\_retx & dl\_bitrate & dl\_err & ul\_err \\
\hline
T0 & 0.3193 & 0.37476 & 0.47826 & 0.66926 & 1.00000 & 0.42875 & 0.955 & 0.00909 & 0.45582 & 0.15981 & 0.340 & 0.00267 & 0 & 0 \\ \hline
T1 & 0.30175 & 0.35564 & 0.45652 & 0.67315 & 0.93333 & 0.43123 & 0.965 & 0.00000 & 0.45630 & 0.15946 & 0.336 & 0.00268 & 0 & 0 \\ \hline
T2 & 0.30175 & 0.36520 & 0.47826 & 0.66926 & 0.93333 & 0.43776 & 0.945 & 0.00260 & 0.46548 & 0.15981 & 0.356 & 0.00265 & 0 & 0 \\
\hline
\end{tabular}}
\end{center}

\textbf{Model Prediction Output:} \texttt{1 (Anomalous)}

\textbf{LIME Local Explanation:}

\begin{center}
\begin{tabular}{|l r|l r|l r|}
\hline
\textbf{Feature} & \textbf{Contrib.} & \textbf{Feature} & \textbf{Contrib.} & \textbf{Feature} & \textbf{Contrib.} \\
\hline
ul\_bitrate\_t-0 $>$ 0.22 & 0.09673 & ul\_retx\_t-0 $>$ 0.30 & 0.07885 & ul\_bitrate\_t-2 $>$ 0.22 & -0.07646 \\\hline
$0.72 <$ dl\_mcs\_t-1 $\leq$ 0.99 & -0.06555 & $0.01 <$ dl\_tx\_t-0 $\leq$ 0.42 & 0.05546 & ul\_retx\_t-1 $>$ 0.30 & 0.05197 \\\hline
$0.14 <$ ul\_tx\_t-0 $\leq$ 0.62 & 0.04699 & epre\_t-0 $\leq$ 0.37 & 0.04515 & $0.01 <$ dl\_tx\_t-1 $\leq$ 0.42 & 0.03832 \\\hline
ul\_retx\_t-2 $>$ 0.30 & 0.03555 & $0.00 <$ dl\_retx\_t-2 $\leq$ 0.01 & 0.03123 & $0.14 <$ ul\_tx\_t-1 $\leq$ 0.62 & 0.02903 \\\hline
dl\_err\_t-1 $\leq$ 0.00 & -0.02899 & epre\_t-2 $\leq$ 0.36 & 0.02518 & ul\_bitrate\_t-1 $>$ 0.22 & -0.02365 \\\hline
$0.14 <$ ul\_tx\_t-2 $\leq$ 0.62 & 0.0234 & $0.01 <$ dl\_tx\_t-2 $\leq$ 0.42 & -0.02197 & ul\_err\_t-0 $\leq$ 0.00 & -0.02156 \\\hline
$0.00 <$ dl\_bitrate\_t-0 $\leq$ 0.06 & 0.01579 & $0.31 <$ p\_ue\_t-2 $\leq$ 0.57 & -0.01194 & $0.31 <$ p\_ue\_t-1 $\leq$ 0.57 & -0.0112 \\\hline
epre\_t-1 $\leq$ 0.36 & 0.01107 & $0.31 <$ p\_ue\_t-0 $\leq$ 0.57 & 0.0108 & dl\_err\_t-2 $\leq$ 0.00 & 0.01077 \\\hline
dl\_retx\_t-1 $\leq$ 0.00 & -0.01061 & pusch\_snr\_t-1 $\leq$ 0.43 & -0.01053 & dl\_retx\_t-0 $\leq$ 0.00 & -0.00867 \\\hline
ul\_err\_t-1 $\leq$ 0.00 & -0.00809 & $0.45 <$ ul\_mcs\_t-0 $\leq$ 0.69 & -0.00651 & $0.00 <$ dl\_bitrate\_t-2 $\leq$ 0.06 & 0.00641 \\\hline
$0.87 <$ cqi\_t-0 $\leq$ 1.00 & -0.00577 & pusch\_snr\_t-0 $\leq$ 0.43 & -0.00525 & $0.00 <$ dl\_bitrate\_t-1 $\leq$ 0.06 & 0.00496 \\\hline
$0.72 <$ dl\_mcs\_t-2 $\leq$ 0.99 & -0.00479 & $0.45 <$ ul\_mcs\_t-1 $\leq$ 0.69 & 0.00422 & $0.87 <$ cqi\_t-1 $\leq$ 1.00 & -0.00378 \\\hline
$0.45 <$ ul\_mcs\_t-2 $\leq$ 0.69 & 0.00343 & dl\_err\_t-0 $\leq$ 0.00 & -0.00265 & $0.72 <$ dl\_mcs\_t-0 $\leq$ 0.99 & 0.00181 \\ \hline
ul\_err\_t-2 $\leq$ 0.00 & -0.001 & pusch\_snr\_t-2 $\leq$ 0.43 & -0.00078 & $0.87 <$ cqi\_t-2 $\leq$ 1.00 & 0.00067 \\
\hline
\end{tabular}
\end{center}

\textbf{SHAP Local Heatmap:}\vspace{-1.5em}
\begin{center}
\resizebox{\textwidth}{!}{%
\begin{tabular}{|l|c| c| c| c| c |c| c| c| c| c |c| c| c |c|}
\hline
\textbf{Timestamp} & epre & pusch\_snr & p\_ue & ul\_mcs & cqi & ul\_bitrate & dl\_mcs & dl\_retx & ul\_tx & dl\_tx & ul\_retx & dl\_bitrate & dl\_err & ul\_err \\
\hline
T0 & 0.00271 & 0.00000 & 0.00000 & 0.00525 & -0.00875 & 0.00000 & -0.00923 & 0.00387 & 0.01334 & 0.06155 & 0.02720 & 0.06972 & 0.00000 & 0.00000 \\
T1 & 0.00000 & 0.00320 & -0.01210 & 0.01151 & 0.00000 & 0.03894 & -0.01946 & 0.00784 & 0.04623 & 0.14729 & 0.03301 & 0.04863 & 0.00000 & 0.00000 \\
T2 & 0.01356 & 0.00000 & 0.00592 & 0.00000 & -0.01502 & 0.15978 & -0.00992 & 0.01900 & 0.14422 & 0.16077 & 0.02937 & 0.02117 & 0.00000 & -0.00522 \\
\hline
\end{tabular}
}
\end{center}
\textbf{Global SHAP Feature Importance:}\vspace{-1.5em}
\begin{center}
\resizebox{\textwidth}{!}{%
\begin{tabular}{|l|c| c| c| c| c |c| c| c| c| c |c| c| c |c|}
\hline
\textbf{Timestamp} & epre & pusch\_snr & p\_ue & ul\_mcs & cqi & ul\_bitrate & dl\_mcs & dl\_retx & ul\_tx & dl\_tx & ul\_retx & dl\_bitrate & dl\_err & ul\_err \\
\hline
T0 & 0.00975 & 0.00844 & 0.00853 & 0.01271 & 0.00466 & 0.01740 & 0.01768 & 0.01208 & 0.01597 & 0.09860 & 0.01823 & 0.03002 & 0.00051 & 0.00514 \\
T1 & 0.00799 & 0.00981 & 0.01421 & 0.00769 & 0.00560 & 0.01767 & 0.03328 & 0.01418 & 0.02568 & 0.07379 & 0.02140 & 0.01703 & 0.00058 & 0.00429 \\
T2 & 0.00549 & 0.00538 & 0.00705 & 0.00464 & 0.00482 & 0.04328 & 0.01549 & 0.01309 & 0.09734 & 0.06278 & 0.01995 & 0.01134 & 0.00053 & 0.00468 \\
\hline
\end{tabular}
}
\end{center}

\textbf{Task For LLM:} \\
Provide a human-readable summary of the model's decision. Highlight which features or patterns most influenced the classification outcome. Assess the likelihood of misclassification and suggest actionable mitigation steps for the network operator.
\end{tcolorbox}
\captionof{figure}{Few-Shot LLM Input Prompt}
\label{fig:llm_input_fs}
\vspace{0.5em}
\twocolumn

\section{Framework Analysis}\label{sec:4}
This section performs an evaluation of the proposed framework. Section~\ref{sec:4.1} describes the dataset and the pre-processing steps. Section~\ref{sec:4.2} details the simulation setup. Finally, Section~\ref{sec:4.4} shows the numerical results of the LSTM performance, XAI module explanations and finally the LLM interpretations.

\begin{table}[!t]
\centering
\caption{\small Overview of DDoS attack types, occurrence times, and participating devices}
\label{tab:dd}
\resizebox{\columnwidth}{!}{%
\begin{tabular}{|c|c|c|c|c|}
\hline 
\textbf{\#} & \textbf{Attack Type} & \textbf{Date}       & \textbf{Time}       & \textbf{Devices (IP)}       \\ \hline \hline
1           & SYN Flood            & 18.08.2024          & 07:00 - 08:00       & \begin{tabular}[c]{@{}c@{}}10.20.10.2 \\ 10.20.10.4\end{tabular} \\ \hline
2           & ICMP Flood           & 19.08.2024          & 07:00 - 09:41       & \begin{tabular}[c]{@{}c@{}}10.20.10.2 \\ 10.20.10.4\end{tabular} \\ \hline
3           & UDP Fragmentation    & 19.08.2024          & 17:00 - 18:00       & \begin{tabular}[c]{@{}c@{}}10.20.10.2 \\ 10.20.10.4\end{tabular} \\ \hline
4           & DNS Flood            & 21.08.2024          & 12:00 - 13:00       & \begin{tabular}[c]{@{}c@{}}10.20.10.2 \\ 10.20.10.4\end{tabular} \\ \hline
5           & GTP-U Flood          & 21.08.2024          & 17:00 - 18:00       & \begin{tabular}[c]{@{}c@{}}10.20.10.2 \\ 10.20.10.4 \\ 10.20.10.6 \\ 10.20.10.8 \\ 10.20.10.10\end{tabular} \\ \hline
\end{tabular}
}
\end{table}

\subsection{Dataset}\label{sec:4.1}
The dataset used in this study comprises raw 5G radio and core metrics obtained from the National Centre of Scientific Research ``Demokritos''~\cite{ncsrd2024}. This dataset was captured from a real-world 5G testbed following 3GPP standards. It documents various DDoS attack scenarios initiated by malicious UEs within a 5G network. The testbed consists of three cells connected to a shared 5G core, with nine UEs in total. Cells 1 and 3, along with the 5G core, are hosted on the Amarisoft Classic platform, while Cell 2 utilizes the Amarisoft Callbox Mini. 
Up to five malicious UEs were chosen to launch different flooding attacks, including SYN, ICMP, UDP Fragmentation, DNS Flood, and GTP-U Flood, as detailed in Table~\ref{tab:dd}. Attack data were recorded over four days, and KPMs were gathered using a data collector interfaced with the 5G network. For each UE, the KPMs include but not limited to downlink/uplink bit rates, channel quality indicators (CQI), total transmitted bytes and transmission errors, sampled every five seconds.
The final dataset consisted of 686,009 KPM reports, including 674,553 benign and 11,456 malicious samples, reflecting a real-world imbalance where attack events are rare (only 1.7\% of all samples). The attack samples were distributed as follows:  SYN flood (1,402 samples, 12. 2\%), ICMP (3,756 samples, 32. 8\%), UDP fragmentation (1,402 samples, 12.2\%), DNS flood (1,399 samples, 12. 2\%) and GTP-U flood (3,497 samples, 30. 5\%).

\begin{table}[!t]
\centering
\caption{\small Comparison of Features Between Normal and Attack Durations, Sorted by Absolute Difference}
\label{tab:feature_comparison}
\begin{tabular}{|l|c|c|c|c|r|}
\hline
\multirow{2}{*}{\textbf{Feature}} & \multicolumn{2}{c|}{\textbf{Normal}} & \multicolumn{2}{c|}{\textbf{Attack}} & \multirow{2}{*}{\textbf{Diff. (\%)}} \\ \cline{2-5}
 & \textbf{Mean} & \textbf{Std} & \textbf{Mean} & \textbf{Std} & \\ \hline \hline
ul\_err        & 0.10          & 0.69          & 2.94         & 10.90        & 2,906.13 \\ \hline
dl\_err        & 0.06          & 0.41          & 0.36         & 1.26         & 546.32  \\ \hline
dl\_bitrate (Mbps)    & 1.16  & 1.57  & 0.05    & 0.04    & -96.05  \\ \hline
ul\_retx       & 56.67         & 74.79         & 106.92       & 86.46        & 88.67   \\ \hline
dl\_tx         & 592.88        & 799.94        & 288.56       & 158.50       & -51.33  \\ \hline
ul\_tx         & 608.60        & 688.99        & 780.64       & 383.15       & 28.27   \\ \hline
dl\_retx       & 35.01         & 57.39         & 41.62        & 48.99        & 18.88   \\ \hline
dl\_mcs        & 19.33         & 7.57          & 15.79        & 9.37         & -18.31  \\ \hline
ul\_bitrate (Mbps)   & 5.32  & 8.44  & 6.29    & 6.83    & 18.15   \\ \hline
cqi            & 12.15         & 3.26          & 10.28        & 3.97         & -15.39  \\ \hline
ul\_mcs        & 18.43         & 7.55          & 16.01        & 7.53         & -13.13  \\ \hline
p\_ue          & -8.78         & 12.07         & -9.80        & 8.02         & -11.62  \\ \hline
pusch\_snr     & 24.77         & 9.08          & 21.93        & 6.67         & -11.47  \\ \hline
epre           & -103.97       & 15.14         & -110.94      & 11.62        & -6.70   \\ \hline
\end{tabular}
\end{table}

\begin{figure}[!t]
    \centering
    \begin{subfigure}[b]{0.48\columnwidth}
        \includegraphics[width=\textwidth]{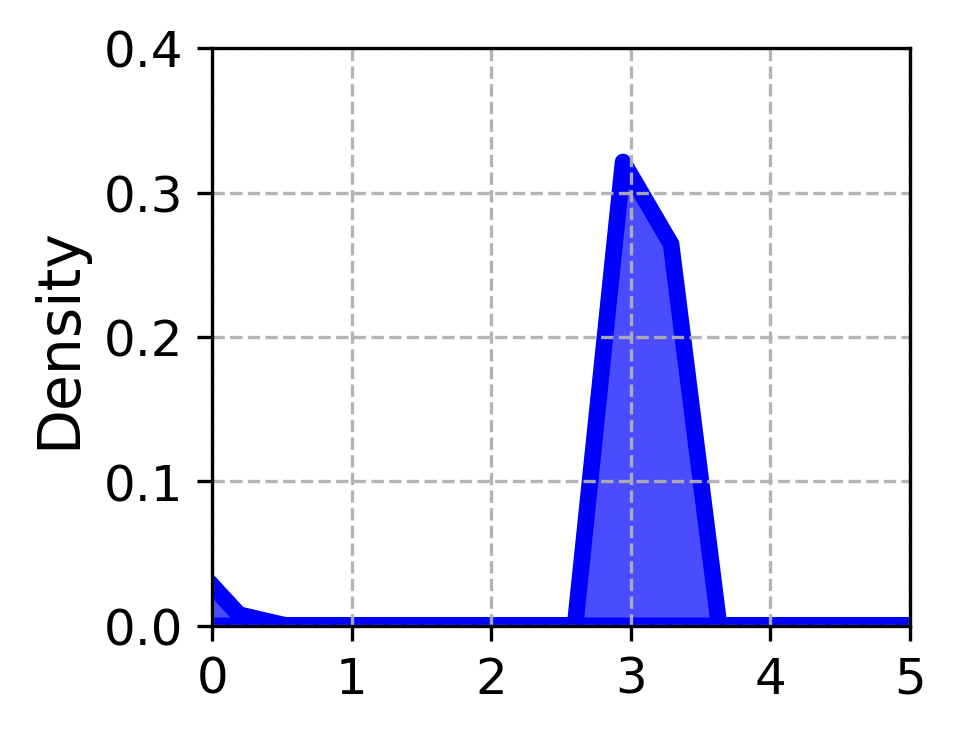}
        \caption{\small Downlink Bitrate - Normal}
        \label{fig:dl_bitrate_attack}
    \end{subfigure}
    \begin{subfigure}[b]{0.48\columnwidth}
        \includegraphics[width=\textwidth]{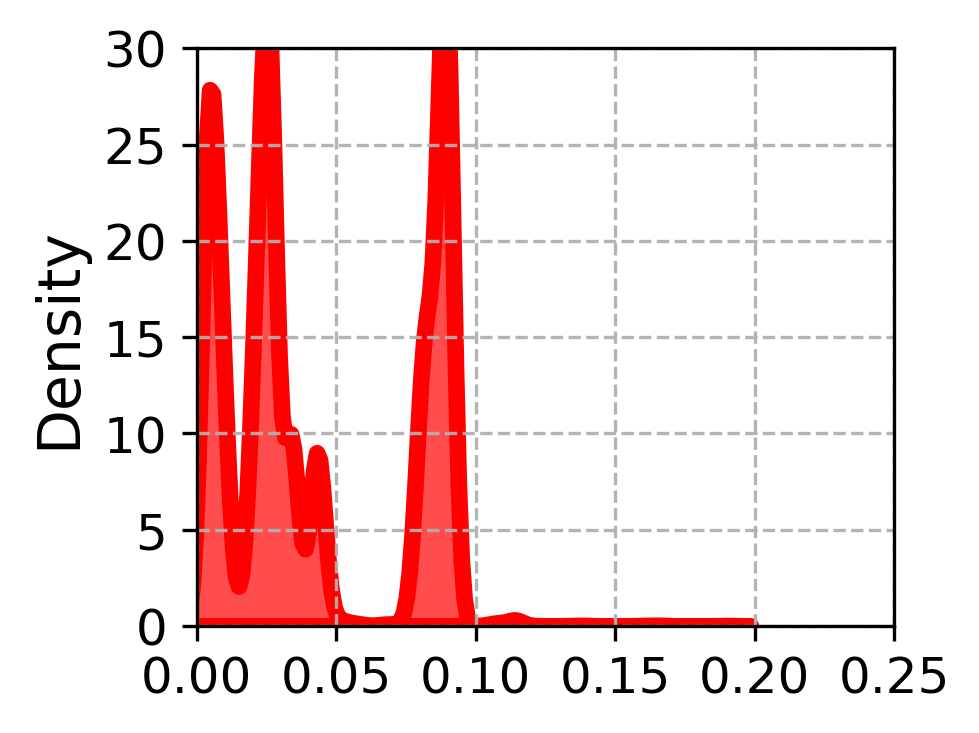}
        \caption{\small Downlink Bitrate - Attack}
        \label{fig:dl_bitrate_normal}
    \end{subfigure}
    \caption{\small KDE Comparison of Downlink Bitrate under Attack and Normal conditions.}
    \label{fig:2by2}
\end{figure}

\begin{figure*}[htbp]
    \centering
    % Row 1: ul_retx
    \begin{subfigure}[b]{0.15\textwidth}
        \includegraphics[width=\textwidth]{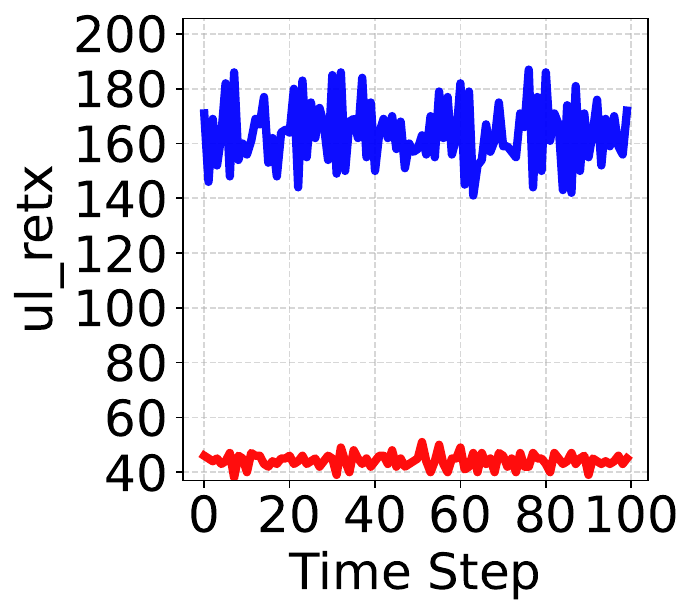}
    \end{subfigure}
    \begin{subfigure}[b]{0.15\textwidth}
        \includegraphics[width=\textwidth]{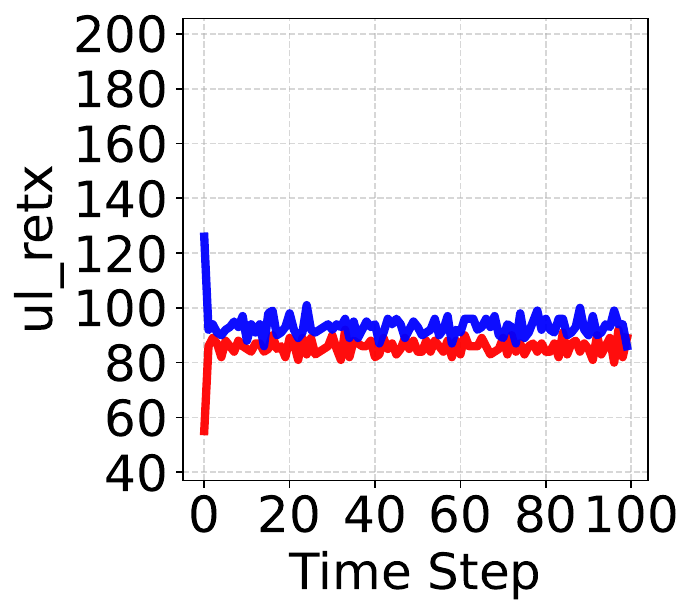}
    \end{subfigure}
    \begin{subfigure}[b]{0.15\textwidth}
        \includegraphics[width=\textwidth]{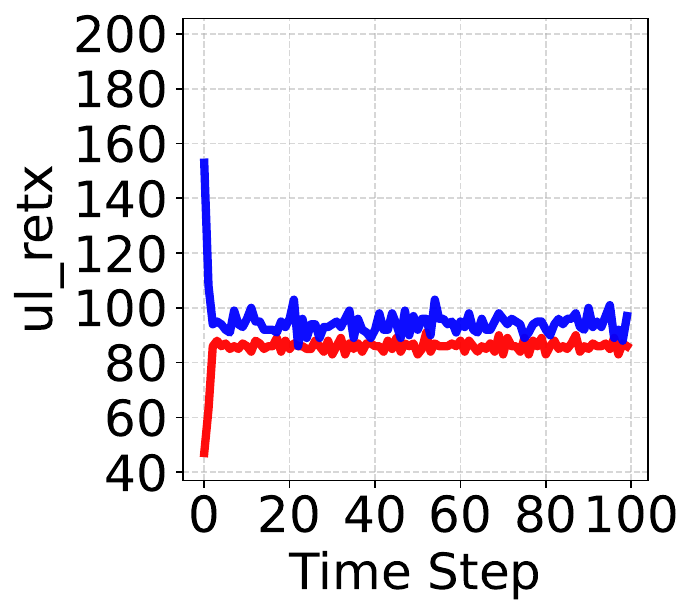}
    \end{subfigure}
    \begin{subfigure}[b]{0.15\textwidth}
        \includegraphics[width=\textwidth]{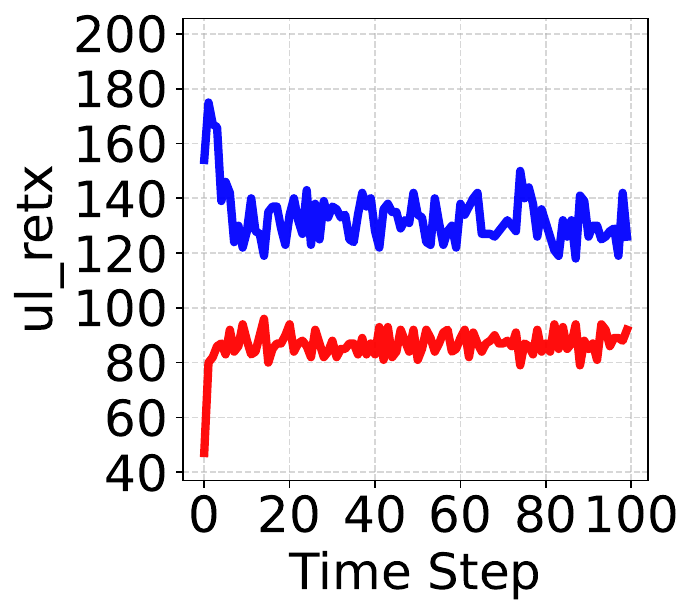}
    \end{subfigure}
    \begin{subfigure}[b]{0.15\textwidth}
        \includegraphics[width=\textwidth]{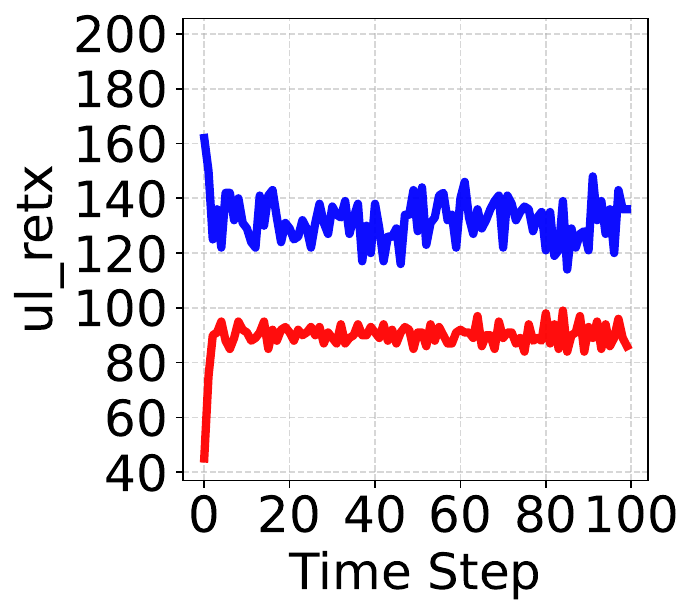}
    \end{subfigure}
    \begin{subfigure}[b]{0.15\textwidth}
        \includegraphics[width=\textwidth]{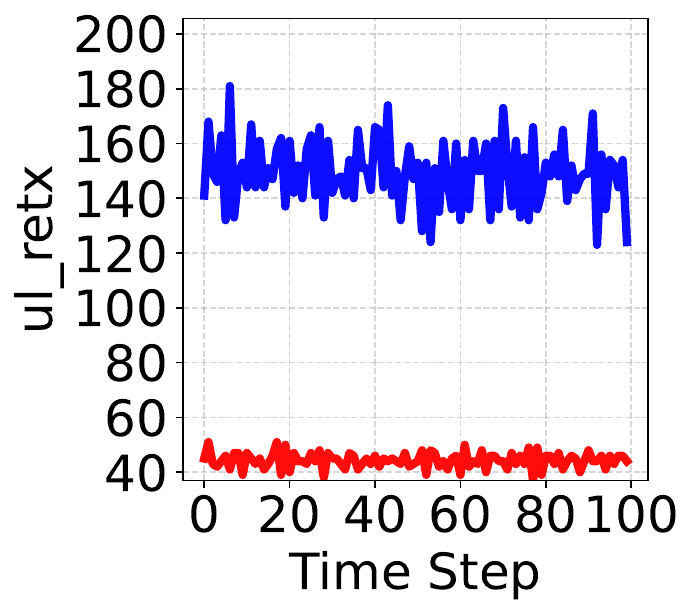}
    \end{subfigure}

    % Row 3: ul_bitrate
    \begin{subfigure}[b]{0.15\textwidth}
        \includegraphics[width=\textwidth]{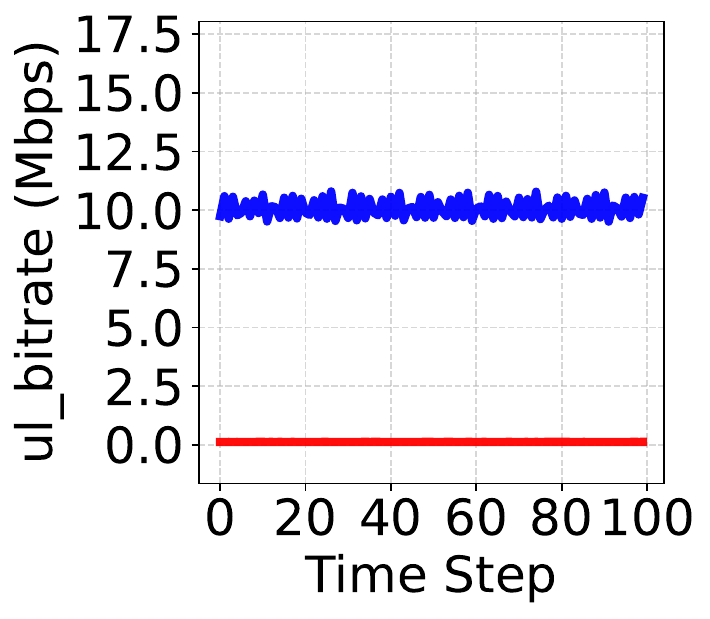}
        \caption{\small Normal}
    \end{subfigure}
    \begin{subfigure}[b]{0.15\textwidth}
        \includegraphics[width=\textwidth]{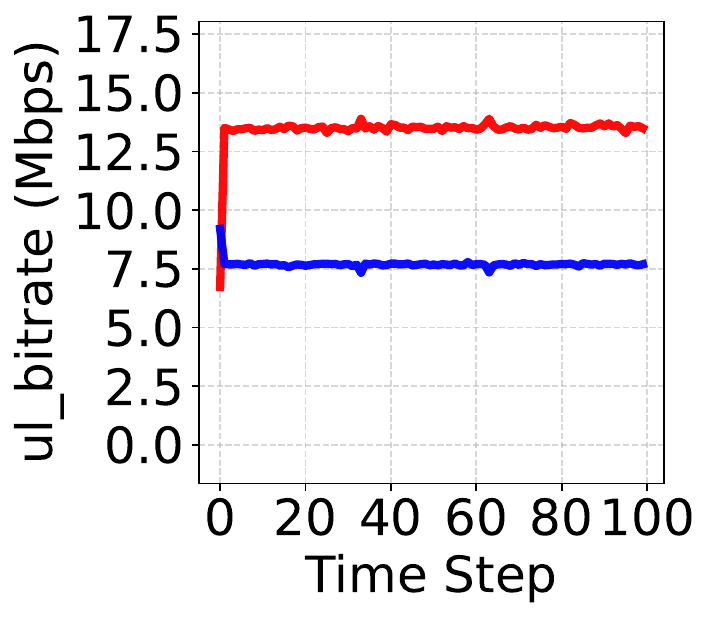}
        \caption{\small SYN}
    \end{subfigure}
    \begin{subfigure}[b]{0.15\textwidth}
        \includegraphics[width=\textwidth]{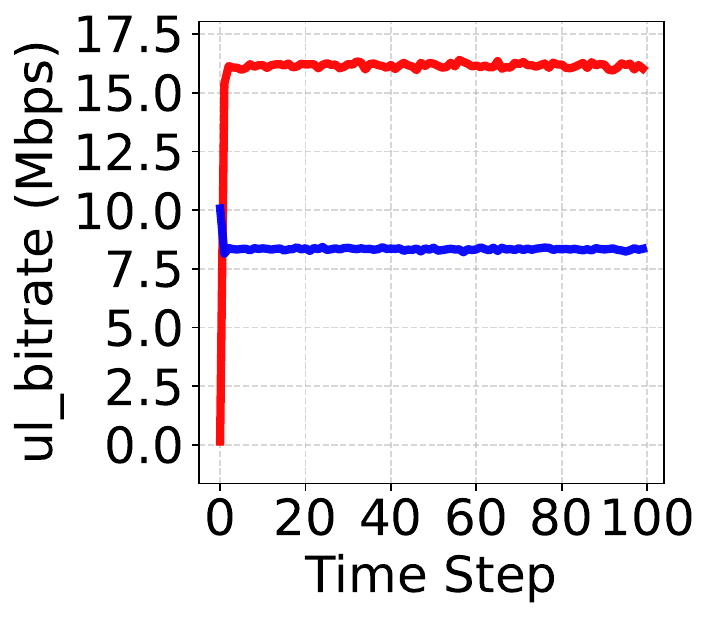}
        \caption{\small ICMP}
    \end{subfigure}
    \begin{subfigure}[b]{0.15\textwidth}
        \includegraphics[width=\textwidth]{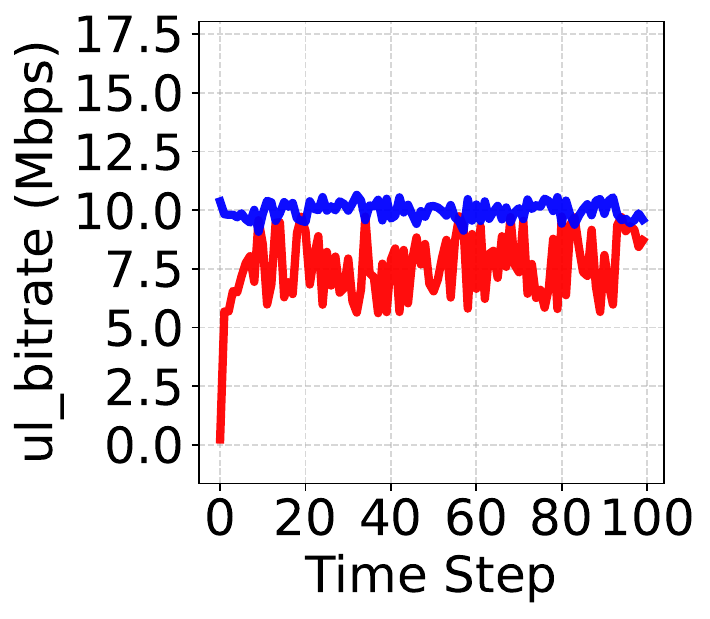}
        \caption{\small UDP}
    \end{subfigure}
    \begin{subfigure}[b]{0.15\textwidth}
        \includegraphics[width=\textwidth]{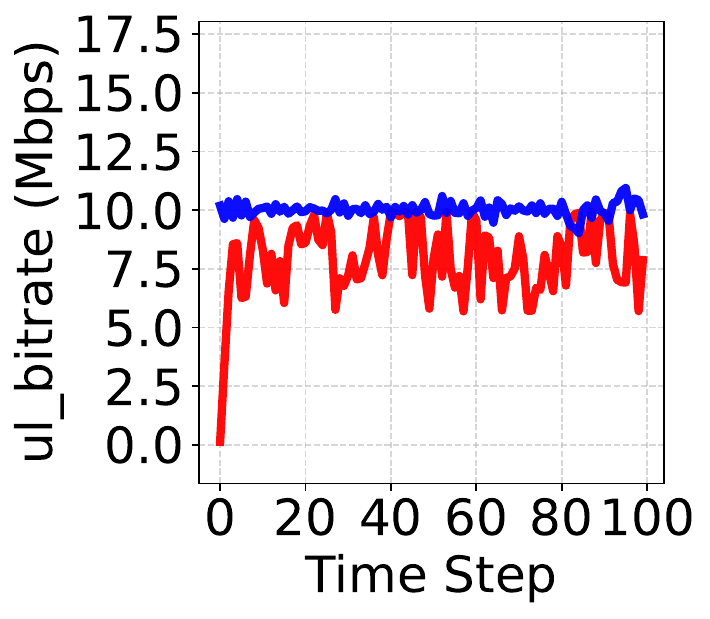}
        \caption{\small DNS}
    \end{subfigure}
    \begin{subfigure}[b]{0.15\textwidth}
        \includegraphics[width=\textwidth]{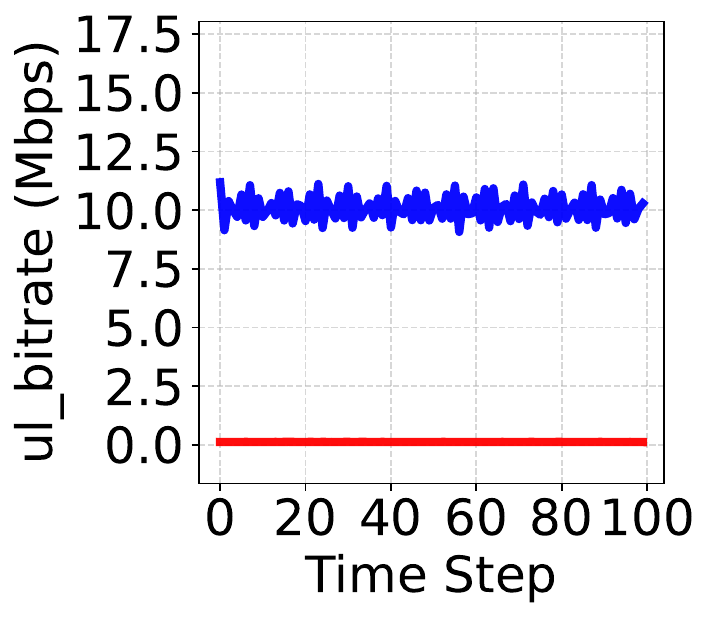}
        \caption{\small GTPU}
    \end{subfigure}
 \caption{\small Feature variation comparison between normal (blue line) and malicious (red line) UEs. Each column corresponds to a different network traffic type: Normal, SYN, ICMP, UDP, DNS, and GTPU, while each row represents a different network feature: \texttt{ul\_retx} (row 1), and \texttt{ul\_bitrate} (row 2). The normal UE data is sampled at the same timestamps as the malicious UE for a direct comparison.}

    \label{fig:comparison_plots}
\end{figure*}

\textbf{Exploratory Data Analysis:} An exploratory data analysis (EDA) was conducted to compare feature distributions during normal and attack periods. This analysis aimed to identify patterns, ranges, and potential indicators of malicious activity. Table~\ref{tab:feature_comparison} highlights the differences in mean values for key features, sorted by their percentage differences. For instance, uplink error rates (\texttt{ul\_err}) and downlink error rates (\texttt{dl\_err}) show a sharp increase during attacks, while downlink bitrates (\texttt{dl\_bitrate}) drop significantly. These insights are supported by visualizations in Fig.~\ref{fig:2by2}, illustrating where data distributions become more skewed under attack conditions. Finally, Fig.~\ref{fig:comparison_plots} illustrates the temporal variation of some features across different attacks, comparing normal (blue) and malicious (red) UEs. The two rows represent different network parameters: uplink retransmissions (\texttt{ul\_retx}), and uplink bitrate (\texttt{ul\_bitrate}), while each column corresponds to a specific traffic type. The normal UE data were aligned with the same timestamps as the malicious UE to provide a direct, time-synchronized comparison. The distinct differences between normal and malicious traffic suggest that these patterns can help in more effective attack detection. 

The dataset was further preprocessed to ensure high-quality input for the LSTM model. Features containing singular values or irrelevant identifiers (e.g., IP addresses) were excluded, while records with NaN values were removed. Data from periods of continuous transmission were selected to accurately represent UE behavior under both normal and attack scenarios.

\subsection{Simulation Setup}\label{sec:4.2}
The proposed model is an LSTM-based neural network designed for sequential network traffic analysis. It consists of an input layer processing sequences of shape $(3,14)$, a 32-unit LSTM layer to extract temporal dependencies, and a final dense layer with sigmoid activation for binary classification.
Extensive fine-tuning experiments were conducted to determine the optimal window size, and past data ratio. The dataset was partitioned into training and test sets using an 80-20 split, with MinMax normalization applied. Training employed the Adam optimizer with binary cross-entropy loss, a batch size of 64, and early stopping (patience = 3). 

Furthermore, we utilized multiple LLMs to generate human-readable summaries and mitigation strategies, selecting models of varying sizes to observe how output quality differs with scale. The models used, in descending order of approximate parameter count, were: \textit{GPT-4-Turbo} ($\sim$1.7T parameters), \textit{DeepSeek-V3-R1} ($\sim$671B), \textit{Mistral-Large-Instruct-2411} ($\sim$123B), and Google’s \textit{Gemini-2.0-Flash} model. All models were accessed programmatically through their respective APIs. When supported, reasoning modes were enabled to assess whether such capabilities improved the quality of the generated outputs.
We quantitatively assessed the outputs of the LLMs using standard readability metrics to evaluate their linguistic clarity and accessibility. Specifically, we used the \texttt{py-readability-metrics} library to compute two representative scores:
(1) Flesch Reading Ease, where higher scores indicate easier-to-read text, and 
(2) Gunning Fog Index, which represents the U.S. school grade level needed to comprehend the text. These metrics provide a quantifiable means of approximating how understandable the LLM-generated content is across models.

\subsection{Results}\label{sec:4.4}
\begin{figure}[!t]
    \centering
    \includegraphics[width=1.00\columnwidth]{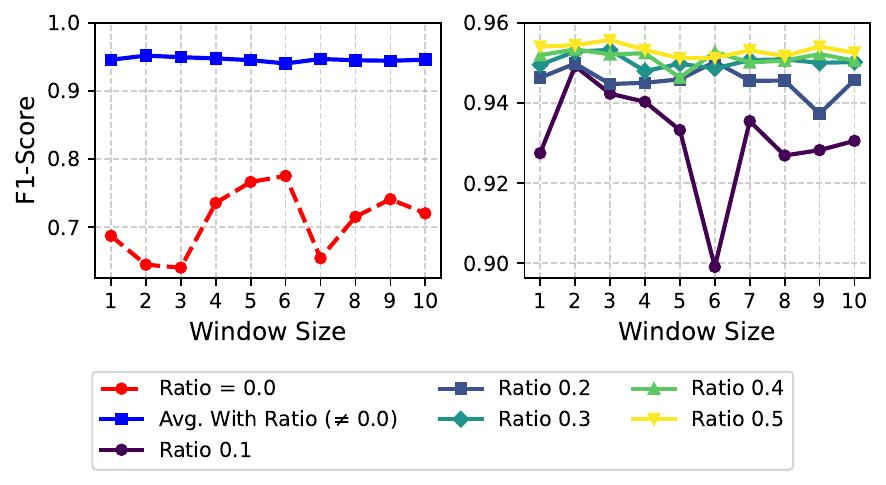}
    \caption{\small Comparison of F1-Score for No Ratio vs. Ratio}
    \label{fig:ratio}
\end{figure}

The first set of experiments evaluated the performance of the LSTM model in attack detection. Specifically, we examined how different combinations of window size and past data ratio influenced the overall model performance while addressing the challenge of catastrophic forgetting, where newly introduced training data overwrites previously acquired knowledge in sequential learning. Fig.~\ref{fig:ratio} illustrates the effect of incorporating different ratios of past training data. The left subplot compares a scenario where no ratio is applied, resulting in an F1-score peaking at 0.80 on average, against one where past knowledge is retained (ratio $\neq 0$), maintaining F1-scores consistently above 0.90. The results indicate that incorporating a ratio stabilizes the model, preventing sharp fluctuations in performance. The right subplot further examines the impact of various ratios ranging from 0.1 to 0.5, showing that while increasing the ratio improves performance, the gains become marginal beyond a certain threshold.

\begin{figure}[!t]
    \centering
    \includegraphics[width=1.00\columnwidth]{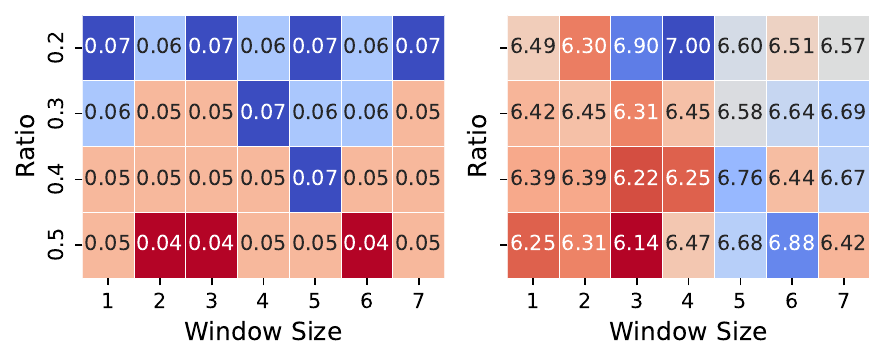}
    \caption{\small Average FPR(\%) (left) and FNR(\%) (right)}
    \label{fig:fprfnr}
\end{figure}

Beyond the effect of ratio selection, we also analyzed the impact of window size and ratio combinations on False Positive Rate (FPR) and False Negative Rate (FNR), as shown in Fig.~\ref{fig:fprfnr}. The heatmap visualization highlights how these hyperparameters influence the trade-off between false alarms (left subplot) and missed detections (right subplot). The results indicate that different window size and ratio combinations lead to variations in FPR and FNR, though the overall differences remain moderate.  Based on these findings, we selected a window size of 3 and a ratio of 0.3 as the optimal configuration for subsequent experiments, as they provided strong performance while keeping training times within a reasonable limit.

Finally, Table~\ref{tab:f1_previous_day} presents the average F1-score of the final LSTM model for each previous day test set, comparing cases where no ratio (0.0) is used versus where a ratio of 0.3 is applied. The results clearly show that incorporating a ratio significantly enhances F1-scores across all previous day evaluations. Without a ratio, the model struggles to retain past knowledge, particularly for Day 4, where performance drops drastically to 0.36. In contrast, applying a ratio of 0.3 effectively preserves past knowledge, leading to consistently high F1-scores, all exceeding 0.96. These findings underscore the importance of balancing new and past training data to maintain model stability and performance, mitigating the risks of catastrophic forgetting.

\begin{table}[!t]
    \centering
    \caption{\small Average F1-score for Previous Days Data with and without Ratio}
    \label{tab:f1_previous_day}
    \begin{tabular}{ccccc}
        \hline
        \textbf{Ratio} & \textbf{Day 1} & \textbf{Day 2} & \textbf{Day 3} & \textbf{Day 4} \\
        \hline\hline
        \textbf{0.0} & 0.69 & 0.64 & 0.70 & 0.36 \\
        \textbf{0.3} & 0.99 & 0.96 & 0.99 & 0.98 \\
        \hline
    \end{tabular}
\end{table}

\begin{table}[!t]
\centering
\caption{\small Performance comparison with previous works using NCSRD.}
\label{tab:comparison_prior_work}
\begin{tabular}{lccc}
\hline
\textbf{Model} & \textbf{F1-Score} & \textbf{FPR (\%)} & \textbf{FNR (\%)} \\
\hline\hline
kNN~\cite{christopoulou_dataset}      & 1.00 & 0.02 & 0.15 \\
XGBoost~\cite{christopoulou_dataset}  & 1.00 & 0.16 & 2.54 \\
CNN~\cite{xylouris_dataset}      & 0.93 & N/A & N/A \\
LSTM~\cite{xylouris_dataset}  & 0.90 & N/A & N/A \\
\textbf{Proposed DDoS Detection}           & 0.98  & 0.05 & 6.31 \\
\hline
\end{tabular}
\end{table}

\begin{figure*}[!t]
    \centering
     \begin{subfigure}[t]{0.24\textwidth}
    \includegraphics[width=\textwidth,height=4cm]{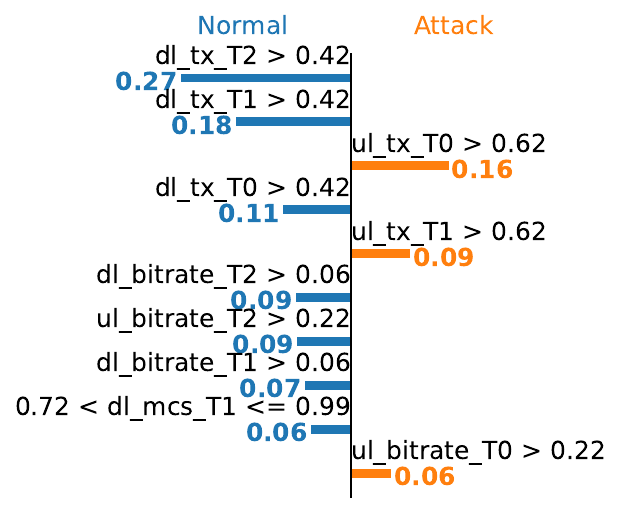}
    \end{subfigure}
    \begin{subfigure}[t]{0.24\textwidth}
        \includegraphics[width=\textwidth,height=4cm]{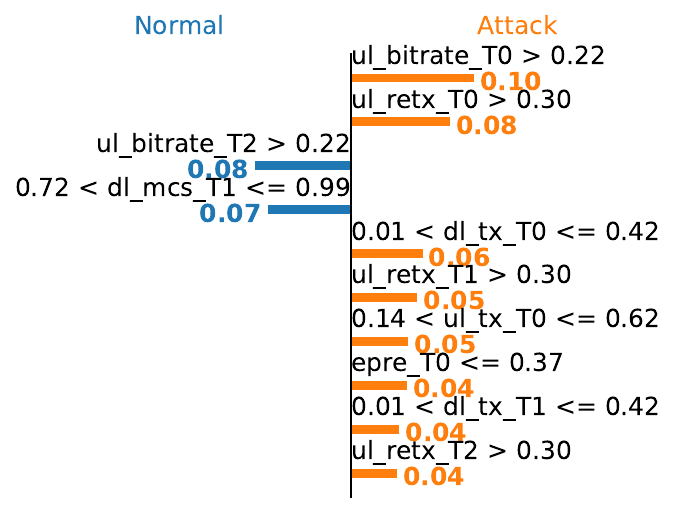}
    \end{subfigure}
    \begin{subfigure}[t]{0.24\textwidth}
       \includegraphics[width=\textwidth,height=4cm]{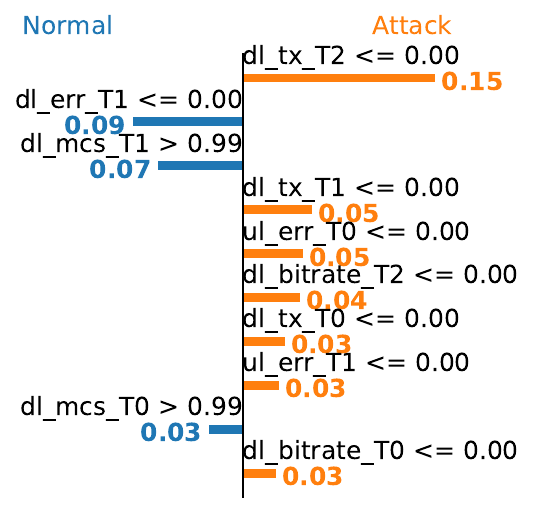}
    \end{subfigure}
    \begin{subfigure}[t]{0.24\textwidth}
       \includegraphics[width=\textwidth,height=4cm]{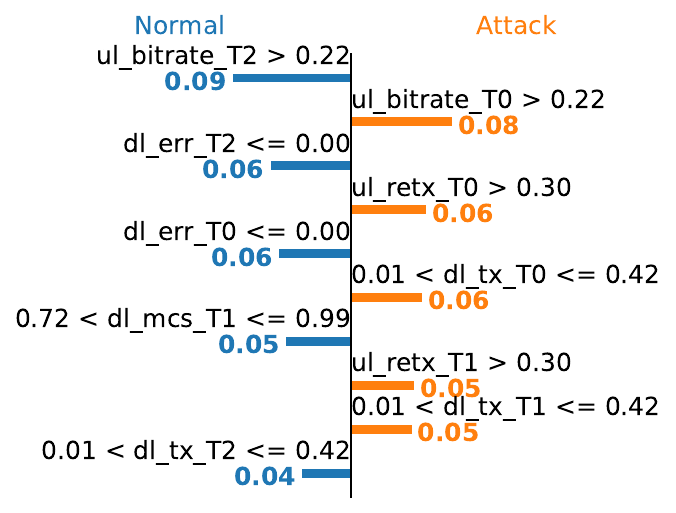}
    \end{subfigure}
    
    \begin{subfigure}[b]{0.24\textwidth}
        \includegraphics[width=\textwidth]{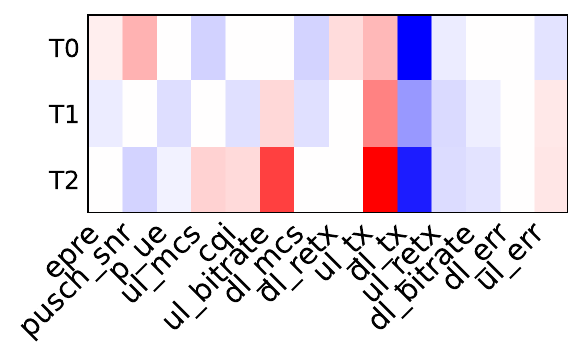}
        \caption{\small TN Instance}
    \end{subfigure}
    \begin{subfigure}[b]{0.24\textwidth}
        \includegraphics[width=\textwidth]{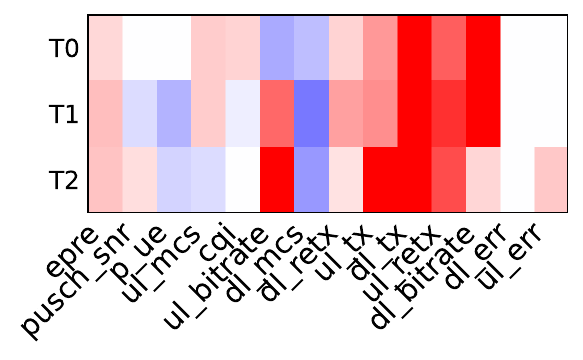}
        \caption{\small TP Instance}
    \end{subfigure}
    \begin{subfigure}[b]{0.24\textwidth}
        \includegraphics[width=\textwidth]{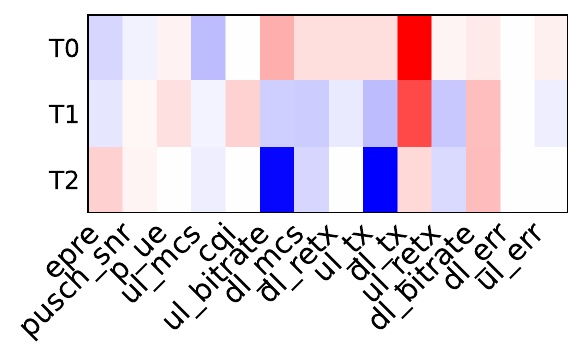}
        \caption{\small FN Instance}
    \end{subfigure}
    \begin{subfigure}[b]{0.24\textwidth}
        \includegraphics[width=\textwidth]{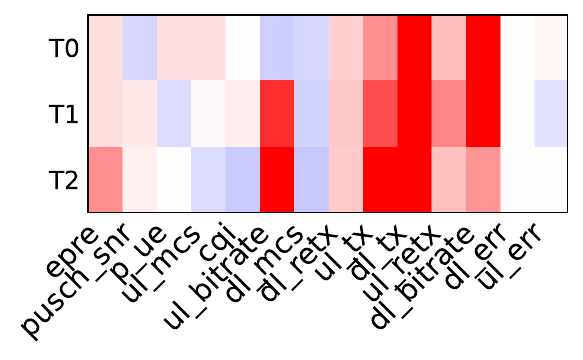}
        \caption{\small FP Instance}
    \end{subfigure}
\caption{\small 
Local explanations for TP, TN, FP, and FN instances. The top row shows the top 10 LIME feature contributions (feature name and value shown; blue: Normal, orange: Attack). The bottom row shows SHAP heatmaps over three time steps (T0–T2), where red supports "Attack" and blue supports "Normal".
}
\label{fig:shap_lime_local}
\end{figure*}

\begin{figure}[!t]
    \centering
    \includegraphics[width=1.00\columnwidth]{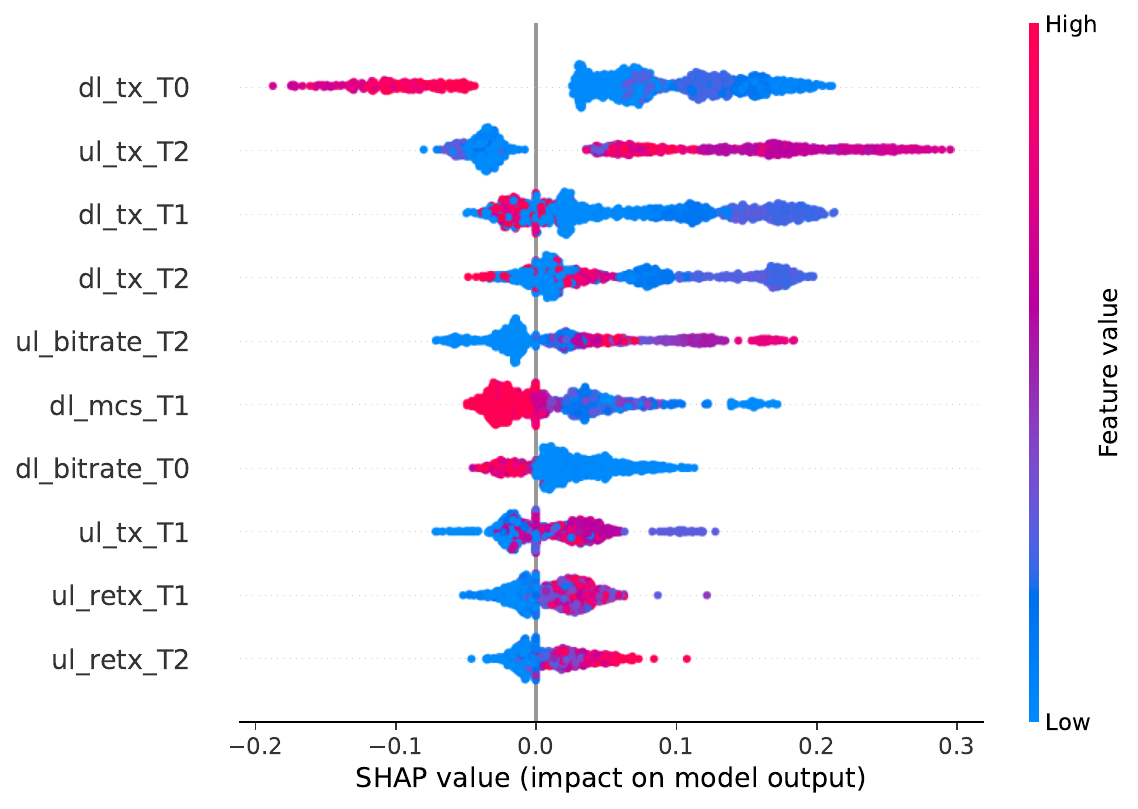}
    \caption{\small Global SHAP beeswarm plot showing the top 10 most influential features across test instances.}
    \label{fig:shap_global_beeswarm}
\end{figure}

Recent studies have evaluated the NCSRD dataset using a range of ML and DL models for DDoS detection. Christopoulou et al.~\cite{christopoulou_dataset} assessed classical ML algorithms such as k-Nearest Neighbors (kNN) and XGBoost, while Xylouris et al.~\cite{xylouris_dataset} explored DL models including CNNs and LSTMs. Table~\ref{tab:comparison_prior_work} summarizes the performance of these approaches alongside our proposed method.

Although the classical ML models achieved strong detection performance~\cite{christopoulou_dataset}, their approach classified each instance independently, without leveraging temporal patterns across sequences, an essential factor for detecting evolving attack behavior. Their top-performing model, kNN with 10 neighbors, also incurs significant inference-time cost, limiting scalability in real-time settings. 
In contrast, our proposed DDoS detection captures temporal dependencies and operates efficiently at inference, making it more suitable for real-time DDoS detection in AI-RAN environments. While Xylouris et al.~\cite{xylouris_dataset} also employed DL architectures capable of modeling temporal information (e.g., LSTM and CNN), differences in model architecture and feature selection contribute to the superior performance of our system.
Our proposed detection system requires an average inference time of 0.03 ms per sample ($\sim$36K FLOPs) on a standard Intel i7-10700 CPU, making it well-suited for near real-time deployment in AI-RAN environments.

Having established the model’s effectiveness, we now focus on explaining its decisions through local and global explanations, using LIME and SHAP to gain insights into individual and aggregated feature contributions. Fig.~\ref{fig:shap_lime_local} presents local explanations for four representative test instances: a true negative (TN), a true positive (TP), a false negative (FN), and a false positive (FP). The top row shows the LIME-based explanations, highlighting the top 10 most influential features for each prediction. These explanations are based on a locally fitted linear model, which interprets each feature's contribution toward the predicted class (either "Normal" or "Attack"). LIME provides human-readable rules, for instance, it may indicate that a feature’s value is greater than, less than, or falls within a specific range. As an example, in the second subplot of the top row, the feature \texttt{dl\_tx} at time step $T0$ contributes to an “Attack” prediction because its value lies between 0.01 and 0.42, as identified by the LIME model.
Furthermore, the second row of Fig.~\ref{fig:shap_lime_local} provides local SHAP explanations of the four test instances. Each heatmap shows the SHAP values across the three time steps (T0-T2) and the 14 input features. Red regions indicate feature-time contributions that increased the likelihood of an “Attack” prediction, while blue regions represent contributions toward the “Normal” class.
In the TP and FP instances, high-intensity red regions are observed for features such as \texttt{ul\_bitrate}, \texttt{ul\_tx}, \texttt{ul\_retx}, \texttt{dl\_tx}, and \texttt{dl\_bitrate}, indicating the model’s strong reliance on these features when predicting an attack. In the TN instance, most feature contributions are of low intensity, except for two strong normal (blue) contributions from \texttt{dl\_tx}, and two strong attack-oriented (red) contributions from \texttt{ul\_bitrate} and \texttt{ul\_tx}. Finally, the FN instance displays a reversed contribution pattern compared to the TN case, where features typically associated with attacks instead contribute toward a normal classification, potentially explaining the misclassification.

\begin{figure}[!t]
    \centering
    \includegraphics[width=1.00\columnwidth]{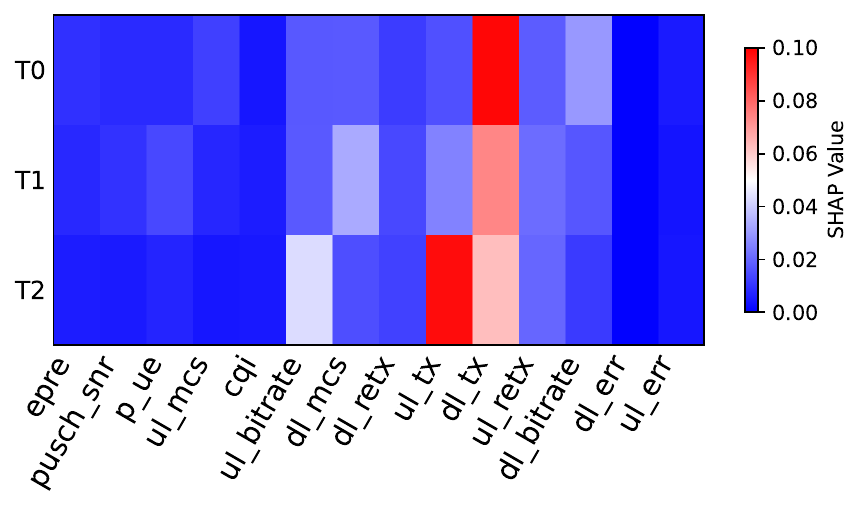}
    \caption{\small Heatmap of the mean absolute SHAP values, showing the overall importance of each feature.}
    \label{fig:shap_global_heatmap}
\end{figure}

Lastly, to understand which input features the LSTM model relies on globally when predicting DDoS, we leveraged the global SHAP explanations. Fig.~\ref{fig:shap_global_beeswarm} presents a beeswarm plot illustrating the top 10 most influential features based on SHAP value magnitudes. Each dot represents a SHAP value for a single instance, colors indicate the actual feature value (from low to high), while the x-axis shows the direction and strength of the feature’s contribution. Features like \texttt{dl\_tx\_T0}, \texttt{ul\_tx\_T2}, and \texttt{ul\_retx\_T1} emerge as highly impactful. To complement this, Fig.~\ref{fig:shap_global_heatmap} provides a heatmap of the mean absolute SHAP values across all test instances, aggregated per feature and time step. This gives a temporal perspective on the importance of each feature. Notably, the most discriminative features cluster around time steps $T1$ and $T2$, with metrics such as \texttt{ul\_tx}, \texttt{dl\_tx}, and \texttt{ul\_bitrate} consistently showing high relevance. 
These global explanation tools highlight the temporal and structural dependencies the model captures when identifying anomalous traffic patterns, which can potentially help in the final interpretations. 

Having analyzed the model's behavior through local and global explanations, we transitioned to the final stage of the proposed framework, leveraging LLMs to transform these technical explanations into human-readable insights. As detailed in Section~\ref{ssec:llm}, we designed special prompts for each test instance we wanted to interpret and then employed a zero-shot prompting strategy to generate informative summaries. The insights generated from a TP instance using zero-shot prompting are shown in Fig.~\ref{fig:llm_output_zs}.
\vspace{0.5em}
\noindent
\begin{tcolorbox}[colback=white, colframe=black, fonttitle=\bfseries, sharp corners, boxrule=0.5pt, breakable, width=\columnwidth]
\label{box:zs_output}
\small
\textit{Here’s a breakdown of the LSTM model's decision, why it classified this sequence as anomalous (likely DDoS), and what actions might help a network operator.}

\textbf{Model Decision:} The LSTM model classified the input as \textbf{anomalous (DDoS attack)} based on temporal network traffic features over 3 time steps.

\textbf{Key Indicators Leading to Anomaly Classification:}
\begin{itemize}
    \item \textbf{High uplink bitrate} at T0 ($\text{ul\_bitrate}_{t0} > 0.22$): Strong LIME contribution ($+0.09664$); confirmed by SHAP at T2.
    \item \textbf{High uplink retransmissions} at T0 and T1 ($\text{ul\_retx} > 0.30$): Suggests congestion or repeated transmissions typical in volumetric attacks.
    \item \textbf{Low epre values} at T0 and T2 ($\text{epre} \leq 0.37$): May indicate abnormal transmission power levels or signal quality manipulation.
    \item \textbf{Asymmetric traffic:} Uplink $\gg$ downlink (e.g., $\text{ul\_tx} \approx 0.45$ vs $\text{dl\_tx} \approx 0.16$).
    \item \textbf{Suppressed downlink activity:} Low $\text{dl\_mcs}$, $\text{dl\_bitrate}$, and $\text{dl\_tx}$.
\end{itemize}

\textbf{Factors Suggesting Potential False Positive:}
\begin{itemize}
    \item \textbf{Good channel quality:} High CQI (e.g., 1.0 at T0 and T1) and $\text{dl\_mcs}$.
    \item \textbf{No transmission errors:} $\text{dl\_err}$ and $\text{ul\_err}$ are all zero.
    \item \textbf{Moderately high uplink bitrate:} Possibly a legitimate upload session.
\end{itemize}

\textbf{Likelihood of Misclassification:} Moderately low ($\sim15\%-25\%$). Traffic resembles a DDoS pattern, but legitimate high-bandwidth uplink is possible.

\textbf{Recommended Actions for Network Operator:}
\begin{itemize}
    \item \textbf{Monitor for continuation:} Observe whether the pattern persists.
    \item \textbf{Cross-reference session data:} IPs, ports, session types.
    \item \textbf{Rate-limit or quarantine:} Throttle or isolate if escalation occurs.
    \item \textbf{Flag for human review:} For audit if later verified as benign.
    \item \textbf{Correlate across users:} Look for broader DDoS coordination patterns.
\end{itemize}

\vspace{4pt}
\textbf{Final Verdict} \\
\textit{The sequence likely represents a true anomaly, exhibiting sustained uplink-heavy behavior and low responsiveness—patterns commonly associated with DDoS activity.}
\end{tcolorbox}\vspace{-0.3em}
\captionof{figure}{Zero-Shot LLM Output}
\label{fig:llm_output_zs} \vspace{0.3em}

The zero-shot prompting strategy used in this stage provided informative and context-aware insights that translated technical model outputs into language suitable for non-expert personnel. These summaries effectively described the rationale for the model's decision, highlighted key contributing characteristics. 
However, we also employed few-shot prompting to refine the LLM-generated outputs, enabling the insights to offer greater detail on the key drivers behind the model’s classification. We presented the LLM with one TP and on TN sample with the appropriate outputs and then asked for the LLM to provide the human-readable insights for a third TP instance as shown in Fig.~\ref{fig:llm_output_fs}.\vspace{0.5em}

\noindent
\begin{tcolorbox}[colback=white, colframe=black, fonttitle=\bfseries, sharp corners, boxrule=0.5pt, breakable, width=\columnwidth]
\label{box:fs_outputs}
\small
The model classified this input as anomalous, driven by strong signs of sustained uplink-dominant traffic and retransmission activity consistent with DDoS patterns:
\begin{itemize}
    \item \textbf{High \texttt{ul\_bitrate}} across all timesteps ($\sim0.43$) — Significantly exceeds the normal mean ($0.0766$) and attack mean ($0.0905$); indicates possible upload flooding.
    \item \textbf{Elevated \texttt{ul\_retx}} ($\sim0.34$) — Well above normal ($0.0864$) and attack ($0.163$) means; suggests frequent retransmissions due to congestion or malicious intent.
    \item \textbf{Asymmetric traffic flow:} $\texttt{ul\_tx} \approx 0.45$–$0.46$ while $\texttt{dl\_tx} \approx 0.159$ — typical of DDoS profiles with minimal downlink response.
    \item \textbf{LIME contributions:}
    \begin{itemize}
        \item $\texttt{ul\_bitrate}_{t0} > 0.22$ ($+0.0967$) and $\texttt{ul\_retx}_{t0} > 0.30$ ($+0.0789$) are top anomaly drivers.
        \item Moderate influence from $\texttt{dl\_tx}$, $\texttt{ul\_tx}$, and low $\texttt{epre}$ (indicative of suboptimal transmission power).
    \end{itemize}
    \item \textbf{SHAP pattern confirmation:}
    \begin{itemize}
        \item Largest positive impact from $\texttt{ul\_bitrate}$ (+0.15978 at T2), followed by $\texttt{ul\_tx}$, $\texttt{ul\_retx}$, and $\texttt{dl\_tx}$.
        \item $\texttt{dl\_bitrate} \approx 0.0027$ — close to the attack mean ($0.0006$), reinforcing uplink-heavy imbalance.
    \end{itemize}
\end{itemize}

\vspace{0.5em}
\textbf{Misclassification Likelihood}
\begin{itemize}
    \item \textbf{Low to Moderate} — Good radio quality (e.g., high $\texttt{cqi}$, $\texttt{dl\_mcs}$) and zero error rates don’t outweigh persistent uplink activity and retransmissions.
    \item Some LIME/SHAP disagreement on $\texttt{ul\_bitrate}_{t2}$ adds uncertainty, but overall indicators support the anomaly label.
\end{itemize}

\vspace{0.5em}
\textbf{Mitigation Steps}
\begin{itemize}
    \item Apply uplink rate-limiting or QoS shaping to reduce risk.
    \item Correlate with session metadata or logs (e.g., NetFlow, PCAP) to confirm traffic legitimacy.
    \item Log and label for future retraining if later deemed benign.
    \item Monitor for propagation — similar traffic across endpoints may signal coordinated DDoS behavior.
\end{itemize}
\end{tcolorbox}
\captionof{figure}{Few-Shot LLM Output}
\label{fig:llm_output_fs} \vspace{0.3em}

Finally, Table~\ref{tab:readability_metrics} highlights the impact of prompting strategies on the readability of LLM-generated explanations. Notably, including reasoning in prompts consistently improved clarity across models, with higher Flesch scores and lower Gunning Fog levels, particularly for OpenAI and DeepSeek, indicating that structured justifications aid interpretability. Few-shot prompting, however, did not universally enhance readability. For some models (e.g., OpenAI), it led to more complex language, reducing accessibility despite additional context. In contrast, models like Mistral and Gemini exhibited moderate gains in reading ease under few-shot conditions, but at the cost of increased syntactic complexity, as reflected in higher Fog indices.

\section{Conclusions}\label{sec:5}
This paper initially presented a survey of existing works on IDS, XAI, and LLM integration for RAN security, highlighting current research efforts and opportunities in this domain. We then introduced an LLM interpretable framework for detecting DDoS attacks in mobile networks, leveraging DL, XAI, and LLMs. Our results demonstrate that an LSTM-based model can reliably detect anomalies in KPMs extracted from UEs with high accuracy. Additionally, integrating SHAP and LIME explanations with LLMs provides interpretable insights accessible to non-technical stakeholders. The inclusion of past data training ratio was also shown to be crucial in mitigating catastrophic forgetting, ensuring stable detection performance over time. 

Future work will explore extending our framework to detect more sophisticated and stealthy attacks (e.g., low-rate or slow DDoS), as well as deploying the proposed xApp and rApp within a real Open RAN emulator environment to assess their operational performance and real-time integration capabilities. We also plan to automate the mitigation pipeline by translating LLM-generated insights into actionable control policies executable via the RIC, enabling a closed-loop detection and response mechanism. Lastly, to further improve the reasoning capabilities of the LLM, we will investigate the use of techniques such as Retrieval Augmented Generation (RAG), which can help the model generate more informed and context-aware mitigation strategies based on external information sources.

\begin{table}[!t]
\centering
\caption{\small Side-by-side comparison of Flesch Reading Ease and Gunning Fog Index scores for each LLM under no-shot and few-shot prompting.}
\label{tab:readability_metrics}
\resizebox{\columnwidth}{!}{%
\begin{tabular}{lllll}
\hline
\textbf{Model} & \multicolumn{2}{c}{\textbf{Flesch Reading Ease}} & \multicolumn{2}{c}{\textbf{Gunning Fog}} \\ \hline\hline
 & No-Shot & Few-Shot & No-Shot & Few-Shot \\
OpenAI & 24.67 & 16.21 & college & college graduate \\
OpenAI (reasoning) & 60.39  & 47.72 & 12 & college \\
DeepSeek & 44.73 &  6.65 & 11 &  college graduate \\
DeepSeek (reasoning) & 48.84 & 51.80 & 10 & 11 \\
Mistral & 37.32 & 45.35 & 12 & college \\
Gemini & 26.34 & 40.13 & college & college \\ \hline
\end{tabular}%
}
\end{table}

\section*{Acknowledgement}
This work has been partially supported by the ORAN-TWIN-X project of CHEDDAR: Communications Hub for Empowering Distributed Cloud Computing Applications and Research, funded by the UK EPSRC under grant numbers EP/Y037421/1 and EP/X040518/1. Also,
it was partially supported in part by the Mobile oRAN for highly Dense Environments (5G MoDE) Project and TUDOR (Towards Ubiquitous 3D Open Resilient Network) Project, two major winners of the U.K.'s Department of Science, Innovation and Technology (DSIT) Open Networks Ecosystem Competition.

\bibliographystyle{IEEEtran}
\bibliography{refereces.bib}
\vspace{-2em}
\begin{IEEEbiography}[{\includegraphics[width=1in,height=1.25in,clip,keepaspectratio]{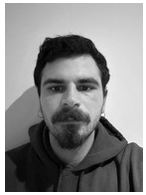}}]{Sotiris Chatzimiltis }
received a B.Sc. degree in computer science from the University of Cyprus, Cyprus, in 2021, and an M.Sc. degree in computer vision, machine learning and robotics from the University of Surrey, U.K., in 2022. He is a PhD student at the Institute for Communication Systems at the University of Surrey. His current research interests include machine learning, intrusion detection systems, distributed AI and Open RAN security.
\end{IEEEbiography}
\vspace{-3em}
\begin{IEEEbiography}
[{\includegraphics[width=1in,height=1in,clip,keepaspectratio]{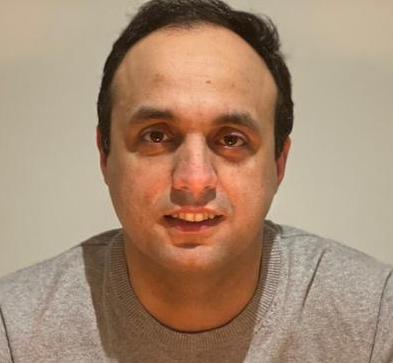}}]{Mohammad Shojafar} \textbf{(M'17-SM'19)} is an Associate Professor in network security and an Intel Innovator, professional ACM member and ACM distinguished speaker, a fellow of the Higher Education Academy, and a Marie Curie alumnus, working in the 5G \& 6G Innovation Centre (5G/6GIC), Institute for Communication Systems (ICS), at the University of Surrey, UK. Before joining 5G/6GIC, he was a senior researcher and a Marie Curie fellow in the SPRITZ Security and Privacy Research group at the University of Padua, Italy. Dr Mohammad secured around $\pounds$1.9M as PI in various EU/UK projects. He is an associate editor in \textit{IEEE Transactions on Network and Service Management}, \textit{IEEE Transactions on Intelligent Transportation Systems}, \textit{IEEE Transactions on Green Communications and Networking}, \textit{IEEE Transactions on Consumer Electronics}, and Computer Networks. %He was also COI of various UK/EU projects like HiPER-RAN (funded by DSIT/UK;2023), APTd5G project (funded by EPSRC/UKI-FNI:2022), ESKMARALD (funded by UK/NCSC;2022), GAUChO, S2C and SAMMClouds (funded by Italian Government;2016-2018). He received his PhD in ICT from Sapienza University of Rome, Rome, Italy, in 2016 with an ``Excellent'' degree. Further information: \url{https://www.surrey.ac.uk/people/mohammad-shojafar}. , including ORAN-TWIN/X (funded by EPSRC/DSIT CHEDDAR Hub UK;2024/2025), D-XPERT (funded by I-UK/UK;2024), 5G MoDE (funded by DSIT/UK;2023), 5G ONE4HDD (funded by DSIT/UK;2023),  TRACE-V2X (funded by EU/MSCA-SE;2023), AUTOTRUST (funded by ESA/EU;2021), PRISENODE (funded by EU/MSCA-IF:2019), and SDN-Sec (funded by Italian Government:2018). 
\end{IEEEbiography}

\begin{IEEEbiography}[{\includegraphics[width=1in,height=1.25in,clip,keepaspectratio]{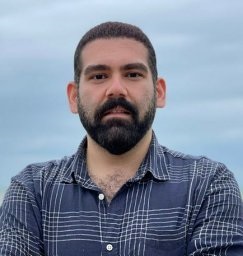}}]{Mahdi Boloursaz Mashhadi}\textbf{(S’14-M’18, SM’23)} is a Lecturer at the 5G/6G Innovation Centre (5G/6GIC) at the Institute for Communication Systems (ICS), University of Surrey (UoS), and a Surrey AI fellow. His research is focused at the intersection of AI/ML with wireless communication, learning and communication co-design, generative AI for telecommunications, and collaborative machine learning. He received B.S., M.S., and Ph.D. degrees in mobile telecommunications from the Sharif University of Technology (SUT), Tehran, Iran. He was a PI/Co-PI for various government and industry funded projects including the UKTIN/DSIT 12M£ national project TUDOR. He is an editor for the Springer Nature Wireless Personal Communications Journal. %He received the Best Paper Award from the IEEE INFOCOM Workshop 2025 and EWDTS 2012, and the Exemplary Reviewer Award from the IEEE ComSoc in 2021 and 2022. He served as a panel judge for the International Telecommunication Union (ITU) on the “AI/ML in 5G” challenge 2021-2022.  He has more than 40 peer reviewed publications and patents in the areas of wireless communications, machine learning, and signal processing.
\end{IEEEbiography}

\begin{IEEEbiography}[{\includegraphics[width=1in,height=1.25in,clip,keepaspectratio]{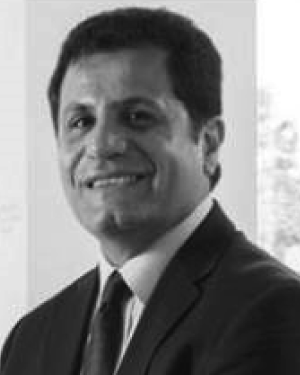}}]{Rahim Tafazolli }\textbf{(Fellow, IEEE)} is currently a Professor in mobile and personal communications and the Director of the Institute of Communication Systems (ICS) and the 5G and 6G Innovation Centre, University of Surrey. He has been active in research for over 30 years and published more than 1200 research papers. He has been a technical advisor to many mobile companies, and has lectured, chaired, and been invited as keynote speaker to a number of IEE and IEEE workshops and conferences. He served as the Chairperson for the EU Expert Group on Mobile Platform (e-mobility SRA), the Chairperson for the Post-IP working Group in e-mobility, and the past Chairperson of WG3 of WWRF. He is nationally and internationally known in the field of mobile communications. In May 2018, he was appointed as a Regius Professor in electronic engineering for recognition of his exceptional contributions to digital communications technologies over the past 30 years. He was elected as a fellow of the U.K. Royal Academy of Engineering, in 2020. He is a fellow of IET and the Wireless World Research Forum (WWRF).
\end{IEEEbiography}
\end{document}